\documentclass[a4paper,11pt]{article}
\pdfoutput=1
\usepackage{jheppub}

\usepackage{epsf}
\usepackage{epsfig}
\usepackage{subfigure}
\usepackage{mathtools}
\usepackage{hhline}
\usepackage{float}
\usepackage{multirow}
\usepackage{nicefrac}
\usepackage{epstopdf}
\usepackage{slashed}
\usepackage{xcolor}

\newcommand{\be}{\begin{eqnarray}}
\newcommand{\ee}{\end{eqnarray}}

\newcommand{\ba}{\begin{array}}
\newcommand{\ea}{\end{array}}
\newcommand{\bee}{\begin{equation}\ba{c}}
\newcommand{\eee}{\ea\end{equation}}

\newcommand{\bi}{\begin{itemize}}
\newcommand{\ei}{\end{itemize}}

\toccontinuoustrue

\title{Contributions of flavor violating couplings of a Higgs boson to $pp\to WW$}
\author{Radovan Derm\'i\v{s}ek,}
\author{Enrico Lunghi}
\author{and Seodong Shin}
\affiliation{Physics Department, Indiana University, Bloomington, IN 47405, USA}
\emailAdd{dermisek@indiana.edu} 
\emailAdd{elunghi@indiana.edu} 
\emailAdd{shinseod@indiana.edu}

\abstract{
We study contributions to $pp\to W^+W^- \to \ell\nu_\ell \ell^\prime\nu_{\ell^\prime}$ in models with a new Higgs boson, $H$, and a neutral lepton, $\nu_{4}$, with couplings $H-\nu_{4}-\nu_{\mu}$ and $W-\nu_{4}-\mu$ through the process $pp \to H \to \nu_4 \nu_\mu \to W \mu \nu_\mu  \to \ell\nu_\ell \mu \nu_\mu$. Contrary to naive expectations, we find that contributions to $pp\to WW$ can be very large while satisfying constraints from standard $H\to WW$ and $H\to\gamma\gamma$ searches. Even the excess observed by ATLAS in $pp\to WW$, if taken at face value, can be easily accommodated. The various kinematic distributions fit nicely the experimentally determined ones. This scenario can arise for example in a two Higgs doublet model with vectorlike leptons.
}
\preprint{
\begin{minipage}{3cm}
\small
\flushright
IUHET-594
\end{minipage}} 

\begin{document}

\maketitle 

\section{Introduction}
\label{sec:introduction}
Recently, the ATLAS collaboration presented a measurement of  $pp \to WW \to \ell \nu_\ell \ell^\prime \nu_{\ell^\prime}$. With $20.3 \; {\rm fb}^{-1}$ of integrated luminosity, the total cross section (including the Higgs contribution) is found to be~\cite{atlasww}
\begin{align}
[\sigma (pp\to WW) + \sigma (gg\to h\to WW^*)]_{\rm exp}= 
71.4^{+1.2}_{-1.2}({\rm stat})^{+5.0}_{-4.4}({\rm syst})^{+2.2}_{-2.1}({\rm lumi}) \;{\rm pb} \; .
\label{eq:atlas}
\end{align}
This result has been obtained by using a next-to-leading order (NLO) Monte Carlo generator (POWHEG ~\cite{Nason:2004rx, Frixione:2007vw, Alioli:2010xd, Melia:2011tj, Nason:2013ydw}) to remove the effects of experimental cuts. The corresponding NLO theoretical prediction in the Standard Model (SM) is~\cite{Campbell:2011bn, atlasww}
\begin{align}
[\sigma (pp\to WW)  + \sigma (gg\to h\to WW^*)]_{\rm th,NLO} =58.7^{+3.0}_{-2.7} \; {\rm pb}
\label{eq:thNLO}
\end{align}
and deviates from the ATLAS result at 2.2 sigma level. Using the same Monte Carlo tools, the CMS collaboration found~\cite{cmswwold}
\begin{align}
\sigma (pp\to WW)_{\rm exp} = 
69.9 \pm 2.8({\rm stat}) \pm 5.6({\rm syst}) \pm 3.1({\rm lumi}) \; {\rm pb}
\label{eq:cmsold}
\end{align}
with $3.5\; {\rm fb}^{-1}$ of integrated luminosity. Taking into account that the SM Higgs contribution is~\cite{Heinemeyer:2013tqa}
\begin{align}
\sigma (gg\to h \to WW^*)_{\rm th} &=4.14^{+7.2\%}_{-7.8\%} \; {\rm pb} \; \label{eq:hww}
\end{align}
the CMS result deviates from the NLO SM prediction at 2 sigma level. Several new physics scenarios have been suggested to explain this excess~\cite{Curtin:2012nn, Jaiswal:2013xra, Rolbiecki:2013fia, Curtin:2013gta, Curtin:2014zua, Kim:2014eva}. 

More recently CMS updated this measurement with $19.4 \; {\rm fb}^{-1}$ of integrated luminosity and found~\cite{CMS:2015uda}
\begin{align}
\sigma (pp\to WW)_{\rm exp}  = 60.1 \pm 0.9({\rm stat}) \pm 3.2({\rm exp}) \pm 3.1({\rm th})  \pm 1.6 ({\rm lumi}) \; {\rm pb} \; .
\label{eq:cms}
\end{align}
This result has been obtained by using the POWHEG NLO Monte Carlo generator but reweighting the simulated $q\bar q\to WW$ events by comparing with a parton level next-to-next-to-leading logarithm (NNLL) calculation in which logarithmic terms that contribute to the $WW$ transverse momentum ($p_T^{WW}$) distribution are resummed~\cite{Meade:2014fca} (see also ref.~\cite{Jaiswal:2014yba} for a discussion of NNLL $p_T^{\rm jet}$ resummation). This resummation mostly affects the calculation of the jet veto efficiency (required to suppress backgrounds from $t\bar t$ and $Wt$ production). Comparing to the NNLO SM prediction~\cite{Gehrmann:2014fva}
\begin{align}
\sigma (pp\to WW)_{\rm th,NNLO} =59.84^{+2.2\%}_{-1.9\%} \; {\rm pb} \; . 
\label{eq:thNNLO} 
\end{align}
CMS finds agreement with the SM. One should point out, however, that the results in eqs.~(\ref{eq:cms}) and (\ref{eq:thNNLO}) have to be compared with care. In fact, fully differential NNLO predictions for $pp\to WW$ are still not available; moreover, the resummation performed in ref.~\cite{Meade:2014fca} is based on NLO matrix elements and considers $p_T^{WW}$ rather than $p_T^{\rm jet}$. 

Note that, the authors of ref.~\cite{Monni:2014zra} find that the NNL effects on the total cross section and on the calculation of the acceptances tend to cancel each other. This suggests that comparing the NLO based ATLAS measurement with the NNLO total cross section can lead to a bias. Depending on which analysis one takes at face value, the deviation of the $pp\to WW$ cross section with respect to the SM expectation is
\begin{align}
\Delta \sigma_{\rm ATLAS} \simeq (13 \pm 6 )\; {\rm pb} 
\quad \text{or} \quad \Delta \sigma_{\rm CMS} \simeq (0 \pm 5 )\; {\rm pb} \; . \label{eq:delta}
\end{align}
This situation will be eventually resolved when better theoretical tools are available leading to an unambiguous interpretation of the experimental results. Given the caveats above, we take the results in eq.~(\ref{eq:delta}) as being compatible with new physics contributions at the 10 pb level either as an explanation of an excess or as a two sigma upper limit. 

In this paper we consider contributions to $pp\to WW \to \ell\nu_\ell \ell^\prime \nu_{\ell^\prime}$ in extensions of the SM that include a new Higgs boson, $H$, and a  neutral lepton, $\nu_{4}$, with couplings $H-\nu_{4}-\nu_{\mu}$ and $W-\nu_{4}-{\mu}$. These new particles can contribute to dilepton final states through following process:
\begin{align}
pp &\to H \to \nu_4 \nu_\mu \to W \mu \nu_\mu  \to \ell\nu_\ell \mu \nu_\mu \; .  \label{eq:nu4}
\end{align}
In order to fix the production cross section, the Higgs boson is assumed to have SM couplings to fermions.  Thus it is dominantly produced in the gluon-gluon fusion channel with the usual SM strength. We further assume that $H$ has no direct coupling to the $W$ boson. This situation arises for example in two Higgs doublet model, in which the light CP-even Higgs boson if fully SM-like in its couplings to gauge boson and thus the heavy CP-even boson has no direct couplings.

The new neutral lepton can originate from extensions of the SM that include new vectorlike lepton families, both SU(2) doublets and singlets. Mixing of vectorlike leptons with the second generation of SM leptons typically implies the appearance of $H-\nu_{4}-\nu_{\mu}$ and $W-\nu_{4}-{\mu}$ couplings and allows for almost arbitrary $H \to \nu_{4}  \nu_{\mu}$ and $\nu_{4} \to W \mu$ branching ratios, see for example the discussion in refs.~\cite{Dermisek:2013gta, Dermisek:2014cia} (where the focus was on the charged lepton sector). 

The signal we consider in eq.~(\ref{eq:nu4}) leads only to $e\mu$ and $\mu\mu$ final states. In order to obtain contributions to the $ee$ mode as well, if desirable, one can introduce an additional neutral lepton that couples exclusively to the the first generation of SM leptons. In fact, simultaneous couplings of one new lepton to both $e$ and $\mu$ leads to unacceptably large contributions to $\mu\to e$ transitions. Given that the cuts required to isolate the $e\mu$ mode are looser than those for the $\mu\mu$ and $ee$ ones (especially to suppress Drell-Yan and $Z\to \ell\ell$ backgrounds) the new physics will mostly affect the statistically dominant $e\mu$ channel.

The main focus of this paper is to show that the models we consider can provide very large contributions to the $pp\to \ell\nu_\ell \ell^\prime \nu_{\ell^\prime}$ even as large as those required by the ATLAS discrepancy in eq. (\ref{eq:delta}). \emph{The crucial reason for this scenario being able to generate such large contributions to the $pp\to \ell\nu_\ell \ell^\prime \nu_{\ell^\prime}$ process is the interplay of a large Higgs production cross section of order 10 pb with the fact that we have only one $W$ that is required to decay leptonically.} For instance, an excess in $WW$ production of about 10 pb corresponds to $\mathcal{O}(0.1)$ pb in the dilepton final state. Having only one $W$ boson in our process implies up to $\mathcal{O}(1)$ pb contribution to the dilepton final state. Therefore we are in an excellent position to explain a significant excess or to place strong constraints in large ranges of masses and branching ratios.

We postpone the discussion of a concrete implementation of these ideas in a complete extension of the SM (e.g. a Two Higgs Doublet Model augmented with vectorlike leptons) to a forthcoming publication~\cite{Dermisek:2015oja}.\footnote{A similar process appears in TeV seesaw neutrino models~\cite{BhupalDev:2012zg}.} The simplified model we consider allows us to present results in a particularly simple way, in terms of the Higgs mass, $m_H$, the neutral heavy lepton mass, $m_{\nu_{4}}$, and the product of branching ratios 
\begin{align}
{\rm BR}(H\to W\ell\nu_\ell) \equiv {\rm BR}(H\to \nu_4 \nu_\mu) \; {\rm BR}(\nu_4\to W \mu)  \;, \label{br}
\end{align}
that can be applied to a variety of specific models.

We require that the simplified model satisfies constraints  from searches for $H\to W^+W^-$~\cite{CMS:bxa, Chatrchyan:2013iaa}  and $H \to \gamma\gamma$~\cite{CMS:2014onr}. These limits apply no matter what the specific realization of the model is. The Higgs boson is assumed to be produced dominantly through the top loop and thus the same loop generates corresponding contribution to  $H \to \gamma\gamma$. Similarly, although we assume no direct coupling of $H$ to $W$, the process in eq.~(\ref{eq:nu4}) contributes to the same final states as $H \to WW$ would. We do not impose any constraints on mass and couplings of $\nu_{4}$ since these would depend on details of the model. For example constraints on pair production of new leptons from searches for anomalous production of multi lepton events, discussed in  ref.~\cite{Dermisek:2014qca}, highly depend on the SU(2) doublet component in the mass eigenstate $\nu_{4}$ and on the mass of extra charged lepton. The results presented in this paper do not depend on these further assumptions. 

The paper is organized as follows. In section~\ref{sec:strategy} we introduce the concept of fiducial cross section and describe our event generation procedure, in section~\ref{sec:constraints} we discuss contributions of our scenario to $pp\to WW$ measurements and the interplay with constraints from the Higgs searches, in section~\ref{sec:distributions} we present actual kinematic distributions for a reference point chosen to fit the nominal excess found by ATLAS and in section~\ref{sec:conclusions} we present our concluding remarks.

\section{Strategy of the analysis}
\label{sec:strategy}
First of all, it is important to realize that the experimental cuts chosen by ATLAS and CMS have different impacts on the SM and on a generic new physics (NP) model. We quantify this statement in terms of acceptances defined as fractions of events that pass a given set of experimental cuts: $A = N_{\rm cuts}/N_{\rm tot}$. The quantity that is constrained by experiments is the so--called \emph{fiducial cross section} defined as the product of the total cross section and the acceptance:
\begin{align}
\sigma^{\rm fid} &=\sigma \; A~.
\label{general_fiducial}
\end{align}
In our case, the fiducial cross sections are given as:
\begin{align}
\sigma^{\rm fid}_{\rm SM} &= \sigma_{pp\to WW}^{\rm SM}  \; {\rm BR}(W \to \ell \nu_{\ell}) \; 
{\rm BR}(W \to \ell^\prime \nu_{\ell^\prime}) \;  A_{\rm SM} \; , \label{sigma_fiducial_SM}\\ 
\sigma^{\rm fid}_{\rm NP} &= \sigma_{pp\to H}^{\rm NP}  \; {\rm BR}(H\to W\ell\nu_\ell) \; 
{\rm BR}(W \to \ell^\prime \nu_{\ell^\prime}) \;  A_{\rm NP} \; , \label{sigma_fiducial_NP}
\end{align}
where ${\rm BR}(H\to W\ell\nu_\ell)$ is defined in eq.~(\ref{br}). The Higgs production cross section $\sigma_{pp\to H}^{\rm NP}$ is assumed to be SM--like and is taken from ref.~\cite{Heinemeyer:2013tqa}. For $e\mu$ final states, $\ell\neq \ell^\prime$, there is an extra combinatoric factor of two in eq.~(\ref{sigma_fiducial_SM}). $A_{\rm SM}$ and $A_{\rm NP}$ are the acceptances corresponding to the ATLAS measurement~\cite{atlasww} and are different for the $\mu\mu$ and $e\mu$ channels. In this paper we focus on the ATLAS analysis~\cite{atlasww} because it explicitly presents results for the fiducial cross sections and allows us to avoid detector simulations.  

In order to allow for an easier interpretation of new physics contributions we define the following effective $pp\to WW$ NP cross sections:
\begin{align}
\sigma_{\rm NP}^{WW} = \frac{\sigma_{\rm NP}^{\rm fid}}{\sigma_{\rm SM}^{\rm fid}} \sigma_{\rm SM}^{WW} = 
\begin{cases}
\frac{\left[ \sigma_{\rm NP}^{\rm fid}\right]_{e\mu}}{2\; {\rm BR} (W\to \ell\nu)^2\; A_{\rm SM}^{e\mu} } \cr
\frac{\left[ \sigma_{\rm NP}^{\rm fid}\right]_{\mu\mu}}{{\rm BR} (W\to \ell\nu)^2 \; A_{\rm SM}^{\mu\mu}  } \cr
\end{cases} \; ,
\end{align}
where the denominators are constants. Although we do not have two $W$ bosons in the final state, this quantity can be directly compared to the ATLAS and CMS results summarized in eq.~(\ref{eq:delta}). Note that the effective cross section $\sigma_{\rm NP}^{\rm WW}$ depends on the ratio of NP and SM acceptances, therefore our results can be directly applied to any future analysis for which this ratio is the same as in the ATLAS study.

The SM acceptances ($A_{\rm SM}$) and the observed fiducial cross sections ($\sigma_{\rm exp}^{\rm fid}$) for the $e\mu$, $ee$ and $\mu\mu$ modes are presented in tables 5 and 9 of ref.~\cite{atlasww}, respectively. The SM fiducial cross sections ($\sigma_{\rm SM}^{\rm fid}$) can be easily obtained using the SM prediction, the $W$ branching ratio into a single lepton flavor ${\rm BR}(W \to \ell \nu) = 0.106$, and the SM acceptances. In table~\ref{table:atlasxsec} we summarize these quantities for each of the $e\mu$, $ee$ and $\mu\mu$ modes. The fiducial cross sections in the $e\mu$ channel are much larger than in the $ee$ and $\mu\mu$ modes because (besides the extra combinatorial factor of 2) the need to reduce Drell-Yan background for same flavor dileptons requires much stronger cuts. The last column of table~\ref{table:atlasxsec} gives the corresponding values of the effective cross section $\sigma_{\rm NP}^{WW}$. 

\begin{table}[t]
\begin{center}
\begin{tabular}{cccccccccc}
\hline
\hline
 & \vphantom{$\Big($}& $\sigma_{\rm exp}^{\rm fid}$ [fb] &&  $\sigma_{\rm SM}^{\rm fid}$ [fb] && $\sigma_{\rm NP}^{\rm fid}$ [fb]  && $\sigma_{\rm NP}^{\rm WW}$ [pb] \\
\hline
$e\mu$ &\vphantom{$\Big($}& $377.8_{-25.6}^{+28.4}$ && $310.6_{-14.3}^{+15.9}$ && $67.2_{-32}^{+30}$ && $12.7^{+6.2}_{-5.8}$ \\
\hline
$ee$ &\vphantom{$\Big($}& $68.5_{-8.0}^{+9.0}$ && $58.6_{-2.7}^{+3.0}$ && $9.9^{+9.4}_{-8.5}$  && $9.9^{+9.5}_{-8.6}$ \\
\hline
$\mu \mu$ &\vphantom{$\Big($}& $74.4_{-7.1}^{+8.1}$ && $63.7_{-2.9}^{+3.3}$ && $10.7_{-7.8}^{+8.6}$  && $9.9^{+8.0}_{-7.3}$ \\
\hline
\hline
\end{tabular}
\caption{Observed ($\sigma_{\rm exp}^{\rm fid}$), expected ($\sigma_{\rm SM}^{\rm fid}$) and required ($\sigma_{\rm NP}^{\rm fid}$) values of the fiducial cross sections for the ATLAS analysis. The last column gives the corresponding required effective cross sections $\sigma_{\rm NP}^{WW}$.}
\label{table:atlasxsec}
\end{center}
\end{table}

For each choice of $m_H$ and $m_{\nu_{4}}$, we extract the acceptance $A_{\rm NP}$ from a Monte Carlo generated event set for the process in eq.~(\ref{eq:nu4}). We use the {\tt MadGraph5}~\cite{Alwall:2014hca} event generator and handle parton shower with {\tt Pythia6}~\cite{Sjostrand:2006za} (implemented in the {\tt MadGraph5 pythia-pgs} package). The resulting {\tt StdHEP} event files are then converted into CERN {\tt root} format using {\tt Delphes}~\cite{deFavereau:2013fsa}. The analysis, that we perform at the shower level, is implemented as a {\tt root} macro. Jet clustering is handled by the {\tt FastJet} package~\cite{Cacciari:2005hq,Cacciari:2011ma}; we adopt an anti-$k_t$ algorithm with $\Delta R = 0.4$. The new physics model has been implemented in {\tt FeynRules}~\cite{Degrande:2011ua}. 

The cuts and the lepton and jet isolation requirements are explicitly given in ref.~\cite{atlasww}. The leading (subleading) leptons in the fiducial region are required to have transverse momentum $p_T > 25$ (20) GeV and pseudo-rapidity in the ranges $|\eta| \in [0,1.37] \cup [1.52,2.47]$ for electrons and $|\eta| < 2.4$ for muons. Dilepton invariant masses, $m_{\ell \ell}$ larger than 10 (15) GeV for the $e\mu$ ($\mu \mu$) channel are required. Additionally, for same flavor leptons, a small window around the $Z$ mass is excluded ($|m_{\ell \ell} - M_Z| > 15$ GeV). Furthermore, cuts on missing transverse and relative transverse momentum are required: $p_T^{\rm miss} \equiv p_T^{\nu + \bar{\nu}} > 20 \; (45) \; {\rm GeV}$ and $p_{T, {\rm Rel}}^{\rm miss} > 15 \; (45) \; {\rm GeV}$ for the $e\mu$ ($\mu\mu$) channel. The latter quantity is defined as $p_{T, {\rm Rel}}^{\rm miss} = p_T^{\rm miss} \times \sin \left(\min \{|\Delta \phi|, \pi/2\} \right)$ where $\Delta \phi$ is the azimuthal angular difference between $\vec{p}_T^{\rm \; miss}$ and the closest jet or lepton in the event. Lepton 4-momenta are corrected by including observed prompt photons emitted within a $\Delta R \leq 0.1$ cone around the lepton direction. Events containing jets with $p_T > 25 \; {\rm GeV}$ and $|\eta| < 4.5$ are vetoed; note that jets within $\Delta R < 0.3$ of a selected electron are not vetoed because they are considered as reconstructed from the electron. In addition to these event selection requirements proper triggers and lepton isolation are also implemented. 

\section{Allowed parameter space and constraints from SM Higgs searches}
\label{sec:constraints}
\begin{table}[t]
\begin{center}
\begin{tabular}{ccccccc}
\hline
\hline
 $\hat m_H$ [GeV] & 120 & 125 & 130 & 160 & 200 & 400  \\
\hline
$e\mu$ & 5.1 fb & 4.8 fb & 4.9 fb & 3.3 fb & 9.7 fb & 3.7 fb \\
\hline
$\mu\mu$ & 5.6 fb & 5.8 fb & 4.5 fb & 3.6 fb & 6.3 fb & 4.0 fb \\
\hline
\hline
\end{tabular}
\caption{The quantities $\beta_{95}^{\cal H}$ for the $e\mu$, $ee$ and $\mu\mu$ channels and for each of the six CMS analyses that we consider (labelled by their $\hat m_H$ value).\label{tab:beta}}
\end{center}
\end{table}
The dominant constraint on $\sigma_{\rm NP}^{\rm fid}$ comes from the $H\to WW$ CMS search presented in refs.~\cite{CMS:bxa, Chatrchyan:2013iaa} where a number of different cuts, each optimized to be sensitive to a SM--like heavy Higgs of a given mass, are considered. For each cut (that we label $\cal H$) CMS, effectively, places a 95\% C.L. upper limit on a fiducial cross section:
\begin{align}
\sigma_{\cal H}^{\rm fid} &= A_{\rm NP}^{\cal H} \; \sigma^{\rm NP} < \beta_{95}^{\cal H} \; ,
\end{align}
where $\sigma^{\rm NP}$ is the same total cross section (including branching ratios) that appears in eq.~(\ref{sigma_fiducial_NP}) and $A_{\rm NP}^{\cal H}$ is the acceptance for the cut selection $\cal H$. Since CMS does not present the results of the analysis in terms of fiducial cross sections, the extraction of these upper limits is not straightforward. We list in table~\ref{tab:beta} the $\beta_{95}^{\cal H}$ that we obtain and relegate the technical details to appendix~\ref{app}. In the table we consider six CMS analyses (labelled by the value $\hat m_H$ of the Higgs mass for which each analysis is optimized) and present separately the $e\mu$ and $\mu\mu$ channels. The implied upper limit on the fiducial cross section (\ref{sigma_fiducial_NP}) is then
\begin{align}
\sigma^{\rm fid}_{\rm NP} &< A_{\rm NP} \; 
\min_\mathcal{H} \left[ \frac{\beta^{\cal H}_{95}}{A_{\rm NP}^{\cal H} } \right] \; .
\label{eq:WWbounds_simp}
\end{align}
Note that our signal, defined in eq.~(\ref{eq:nu4}), and the SM Higgs decay ($H\to WW\to \mu\nu_\mu \ell\nu_{\ell}$) are topologically different. The cuts adopted in refs.~\cite{CMS:bxa, Chatrchyan:2013iaa}, not being optimized to our signal, result in $H\to WW$ acceptances $A_{\rm NP}^{\cal H}$ for both $e\mu$ and $\mu\mu$ final states in the range $[0.01,4]\%$ when varying $m_{\nu_{4}} \in [120,250] \; {\rm GeV}$ and $m_H \in [130,250] \; {\rm GeV}$. The $pp\to WW$ acceptances that we find for the $e\mu$ and $\mu\mu$ cases are in the ranges $[A_{\rm NP}]_{e\mu} \in [24,29]\%$ and $[A_{\rm NP}]_{\mu\mu} \in [5,15]\%$. Therefore, we find that the ratio of acceptances $A_{\rm NP}/A_{\rm NP}^{\cal H}$ has a very strong dependence on the $H$ and $\nu_{4}$ masses and can be very large. This, in turn, implies that, in some regions of the $[m_H,m_{\nu_{4}}]$ plane, we can fully explain the $WW$ excesses summarized in table~\ref{table:atlasxsec} while simultaneously satisfying the limits from $H\to WW$ searches.

\begin{figure}
\begin{center}
\includegraphics[width=.495\linewidth]{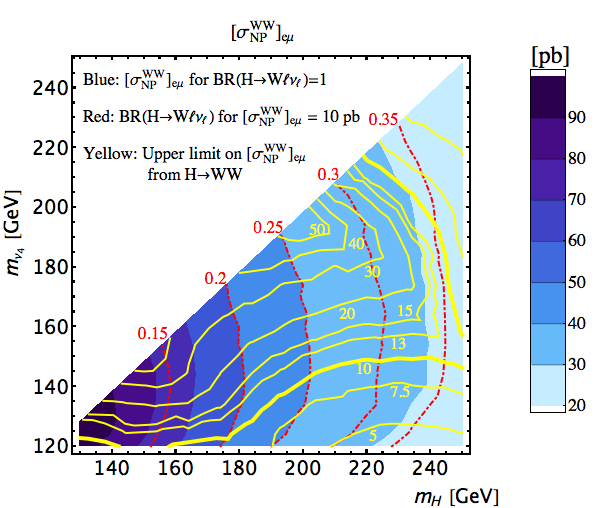}
\includegraphics[width=.495\linewidth]{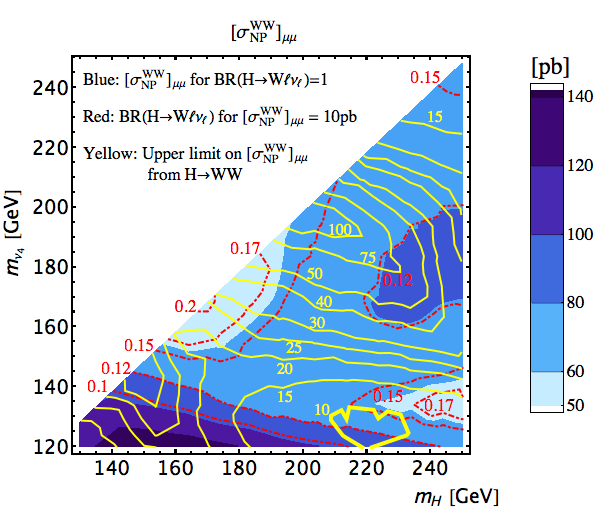}
\caption{In shades of blue we plot contours of $\sigma_{\rm NP}^{\rm WW}$ taking ${\rm BR} (H\to W\ell\nu_\ell) = 1$ for the $e\mu$ (left panel) and $\mu\mu$ (right panel) final states. The dot-dashed red contours are the values of the branching ratios ${\rm BR} (H\to W\ell\nu_\ell)$ required to obtain reference values $\sigma_{\rm NP}^{\rm WW} = 10\; {\rm pb}$ for both the $e\mu$ and $\mu\mu$ modes. The yellow contours are the 95\% C.L. upper limits on the effective cross sections implied by the CMS $H\to WW$ searches summarized in eq.~(\ref{eq:WWbounds_simp}). Branching ratios corresponding to different values of $\sigma_{\rm NP}^{\rm WW}$ can be easily obtained by rescaling.
\label{fig:MhMnu}
}
\end{center}
\end{figure}
The main results of our study are presented in figure~\ref{fig:MhMnu}. In shades of blue we plot contours of $\sigma_{\rm NP}^{\rm WW}$ taking ${\rm BR} (H\to W\ell\nu_\ell) = 1$. We also plot, with dot-dashed red contours, the values of the branching ratios ${\rm BR} (H\to W\ell\nu_\ell)$ required to obtain reference values $\sigma_{\rm NP}^{\rm WW} = 10\; {\rm pb}$ for both the $e\mu$ and $\mu\mu$ modes. The yellow contours in figure~\ref{fig:MhMnu} are the 95\% C.L. upper limits on $\sigma_{\rm NP}^{\rm WW}$ implied by the CMS $H\to WW$ searches summarized in eq.~(\ref{eq:WWbounds_simp}). 

Note that since the acceptance $[A_{\rm NP}]_{e\mu}$ is approximatively constant (in the region we consider in figure~\ref{fig:MhMnu}), the effective cross section contours are essentially controlled by the $pp\to H$ cross section. The cuts adopted in ref.~\cite{atlasww} for the $\mu\mu$ final state are much stronger than those for the $e\mu$ channel. The acceptances $[A_{\rm NP}]_{\mu\mu}$ are, therefore, smaller than $[A_{\rm NP}]_{e\mu}$ and display a more marked dependence on the $H$ and $\nu_{4}$ masses. The structure of the contours we find is correspondingly more complicated. 

The shape of the $H\to WW$ constraints (yellow contours) can be understood as follows. The energy of the SM neutrino produced in the heavy Higgs decay $H\to\nu_{4} \nu_\ell$ scales as $(m_H^2-m_{\nu_{4}}^2)/2m_H$, therefore in the bottom right regions of the $[m_H,m_{\nu_{4}}]$ plane it tends to be very large. This implies a much larger missing energy $E_T^{\rm miss}$ and, in turn, a larger value for the transverse mass $m_T = \sqrt{2 p_T^{\ell \ell} E_T^{\rm miss} ( 1- \cos \Delta \phi_{E_T^{\rm miss}, \ell \ell}})$, where $p_T^{\ell\ell}$ is the transverse momentum of the dilepton system and $\Delta \phi_{E_T^{\rm miss}, \ell \ell}$ is the azimuthal angle difference between $\vec E_T^{\rm miss}$ and $\vec p_T^{\; \ell\ell}$. Since $H\to WW$ searches focus on large $m_T$ values, points in this region have larger $A_{\rm NP}^{\cal H}$ acceptances and are more easily excluded. A similar argument explains why regions at large $m_{\nu_{4}}$ tend to be excluded. Both charged leptons appear in the $\nu_{4}$ decay chain, therefore larger heavy neutral lepton masses imply larger $p_T^{\ell\ell}$ and, in turn, larger $m_T$, strengthening the impact of the $H\to WW$ constraint. 

In large ranges of masses and branching ratios experimental limits on $\sigma_{\rm NP}^{\rm WW}$ imply stronger constraints than direct searches for heavy Higgses decaying to $WW$. For instance, let us focus on $\sigma_{\rm NP}^{\rm WW} = 10\; {\rm pb}$ in the $e\mu$ channel that can be considered either as approximatively the value of the ATLAS excess or the two sigma upper limit implied by the CMS result. In the region of the $[m_H,m_{\nu_4}]$ plane below the the 10 pb yellow contour such large contributions to the effective cross section are excluded at 95\% C.L. by the $H\to WW$ searches. In the region above the contour the values of ${\rm BR}(H\to W\ell\nu)$ required to get exactly 10 pb are given by the red labels. If we require $\sigma_{\rm NP}^{\rm WW}< 10 \; {\rm pb}$, the red labels become upper limits on ${\rm BR}(H\to W\ell\nu)$. The values of branching ratios corresponding to different effective cross sections can be easily obtained by simple rescaling. 

\bigskip

\begin{figure}[t]
\begin{center}
\includegraphics[width=.6\linewidth]{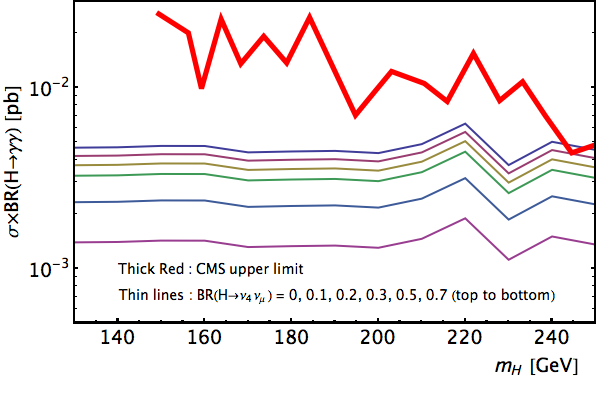}
\caption{Predicted values of $\sigma(pp\to H) \times {\rm BR}(H\to \gamma\gamma)$ as functions of the Higgs mass for different values of the $H\to \nu_{4} \nu_\mu$ branching ratio. The CMS limit is shown with a thick red line. \label{fig:diphoton}}
\end{center}
\end{figure}
Let us now discuss the bounds that we obtain from searches for heavy Higgs bosons decaying to two photons~\cite{CMS:2014onr}. In our simplified model independent analysis we work under the assumption that the light Higgs is purely SM--like and that the heavy Higgs $H$ has no direct coupling to the $W$ boson. This implies that $H\to\gamma\gamma$ partial width we find in our scenario is controlled by the top loop and is, therefore, much smaller than the corresponding partial width for SM--like heavy Higgses. For instance, for $m_H = 155 \; {\rm GeV}$ this effect suppresses the $H\to\gamma\gamma$ partial width by a factor $5\times 10^{-2}$.

The limits presented in ref.~\cite{CMS:2014onr} constrain the product $\sigma(pp\to H) \times {\rm BR}(H\to \gamma\gamma)$. In our scenario the Higgs  production cross section is fixed to its SM--like value~\cite{Heinemeyer:2013tqa} while the branching ratio into diphoton is affected only by impact of $\Gamma (H\to \nu_{4} \nu_\mu)$ on the total width:
\begin{align}
\sigma(pp\to H) \frac{\Gamma_{H\gamma\gamma}}{\Gamma_{Hbb}+\Gamma_{Hgg}+\Gamma_{H\gamma\gamma}+\Gamma_{H\nu_{4}\nu_\mu}} < 
\Big[ \sigma(pp\to H) {\rm  BR} (H\to \gamma\gamma) \Big]_{\rm exp} \; .
\end{align}
In figure~\ref{fig:diphoton} we show predicted values of $\sigma(pp\to H) \times {\rm BR}(H\to \gamma\gamma)$ as functions of the Higgs mass for different values of the $H\to \nu_{4} \nu_\mu$ branching ratio. The CMS limit is shown with a thick red line. We see that present experimental bounds do not constrain our model.

\section{Explanation of the ATLAS excess and kinematic distributions}
\label{sec:distributions}
Let us now consider the excesses observed by ATLAS at face value with the understanding that NLL effects (as hinted by the CMS study) might reduce it sizably. In order to fully account for the observed excesses in the $e\mu$, $\mu\mu$ and $ee$ channels while avoiding lepton flavor violation constraints, we introduce an additional heavy neutral lepton, $\nu_5$, that couples exclusively to the first generation of SM leptons. For simplicity, we assume that the couplings $H-\nu_5-e$ and $W-\nu_5-e$ and the mass of $\nu_5$ are identical to those of $\nu_4$. The processes we consider are:
\begin{align}
pp &\to H \to \nu_4 \nu_\mu \to W \mu \nu_\mu  \to \ell\nu_\ell \mu \nu_\mu \; ,  \label{eq:nu4NEW}\\
pp &\to H \to \nu_5 \nu_e \to W e \nu_e \;\, \to \ell\nu_\ell e \nu_e \; . \label{eq:nu5NEW}
\end{align}
The results for this scenario can be directly obtained from those presented in figure~\ref{fig:MhMnu}. The only difference is that for the $e\mu$ case the branching ratio ${\rm BR} (H\to W\ell\nu)$ has to be interpreted as
\begin{align}
{\rm BR}(H\to W\ell\nu_\ell) &\equiv {\rm BR}(H\to \nu_4 \nu_\mu) \; {\rm BR}(\nu_4\to W \mu) + {\rm BR}(H\to \nu_5 \nu_e) \; {\rm BR}(\nu_5\to W e)  \;, \label{br45} 
\end{align}
while for the $\mu\mu$ case it remains the same as in eq.~(\ref{br}) and for the $ee$ case it is ${\rm BR}(H\to \nu_5 \nu_e) \; {\rm BR}(\nu_5\to W e)$. As we pointed out above, we see that in a large region of masses and branching ratios we can easily explain the excess observed by ATLAS while satisfying constraints from $H\to WW$. In particular these are the regions above the yellow contours corresponding to the central values given in the last column of table~\ref{table:atlasxsec} (about 13 and 10 pb for the $e\mu$ and $ee/\mu\mu$ cases).  

The next important step is to check whether the kinematic distributions that we obtain for points in the allowed regions match the observed ones. In ref.~\cite{atlasww} the following seven quantities are considered: transverse momentum, $p_T$, of the leading and subleading lepton; the azimuthal angle, $\Delta \phi_{\ell\ell}$, transverse momentum, $p_T(\ell\ell)$, and invariant mass, $m_{\ell\ell}$, of the dilepton system; the transverse mass, $m_T(\ell\ell+E_T^{\rm miss})$, and the transverse momentum, $p_T(\ell\ell+E_T^{\rm miss})$, of the dilepton plus missing transverse energy system.
\begin{figure}
\begin{center}
\includegraphics[width=.49\linewidth]{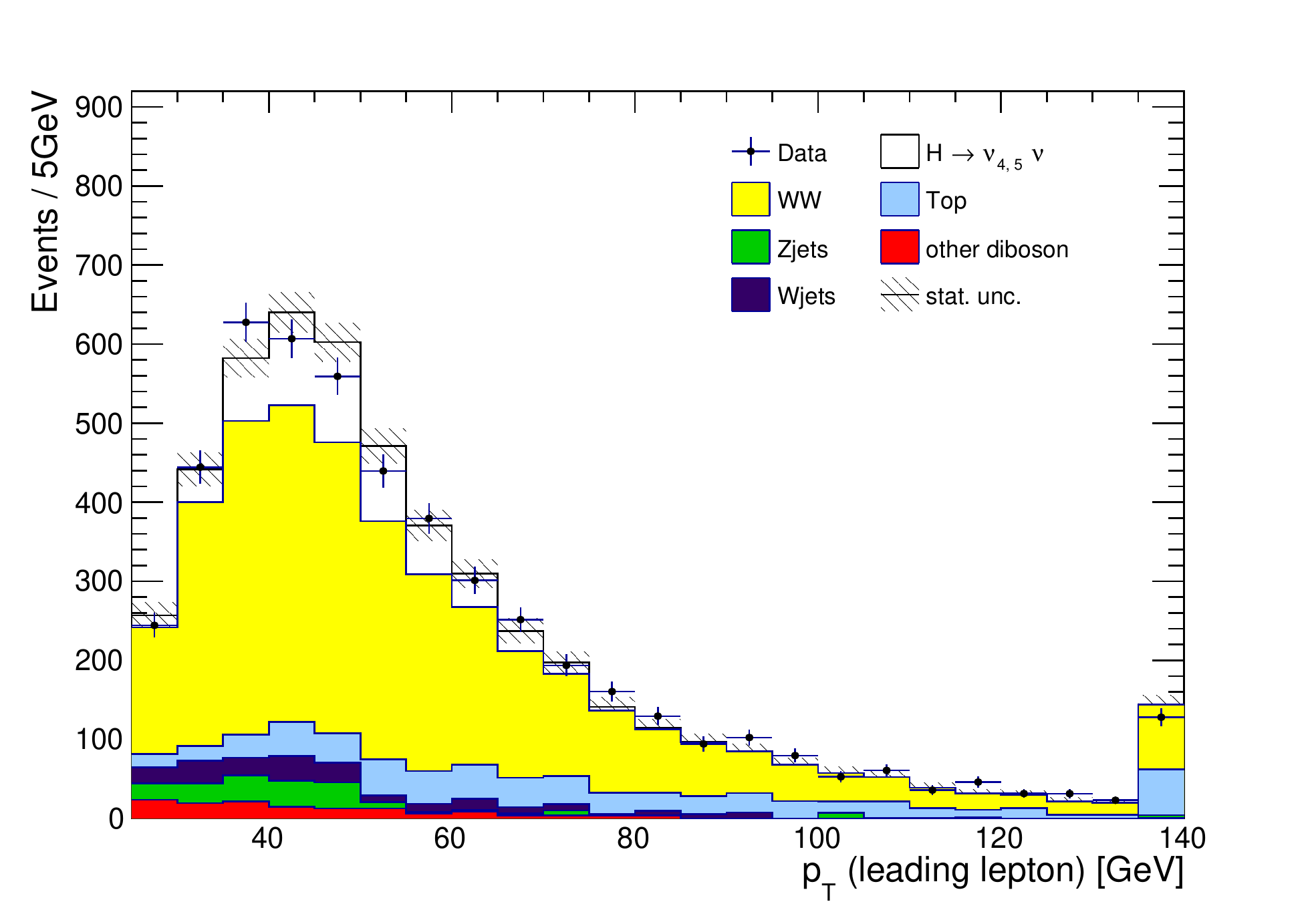}
\includegraphics[width=.49\linewidth]{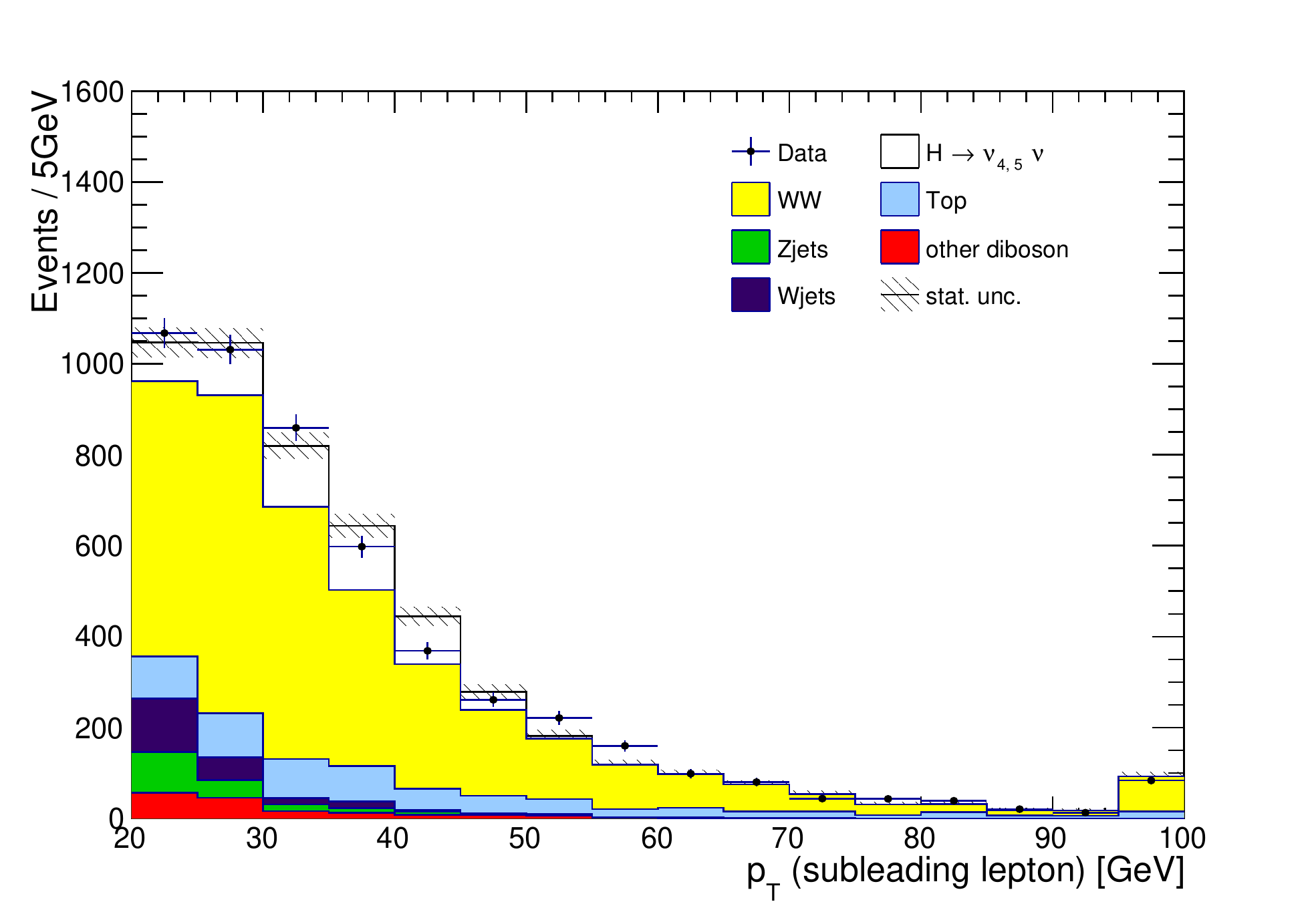}
\includegraphics[width=.49\linewidth]{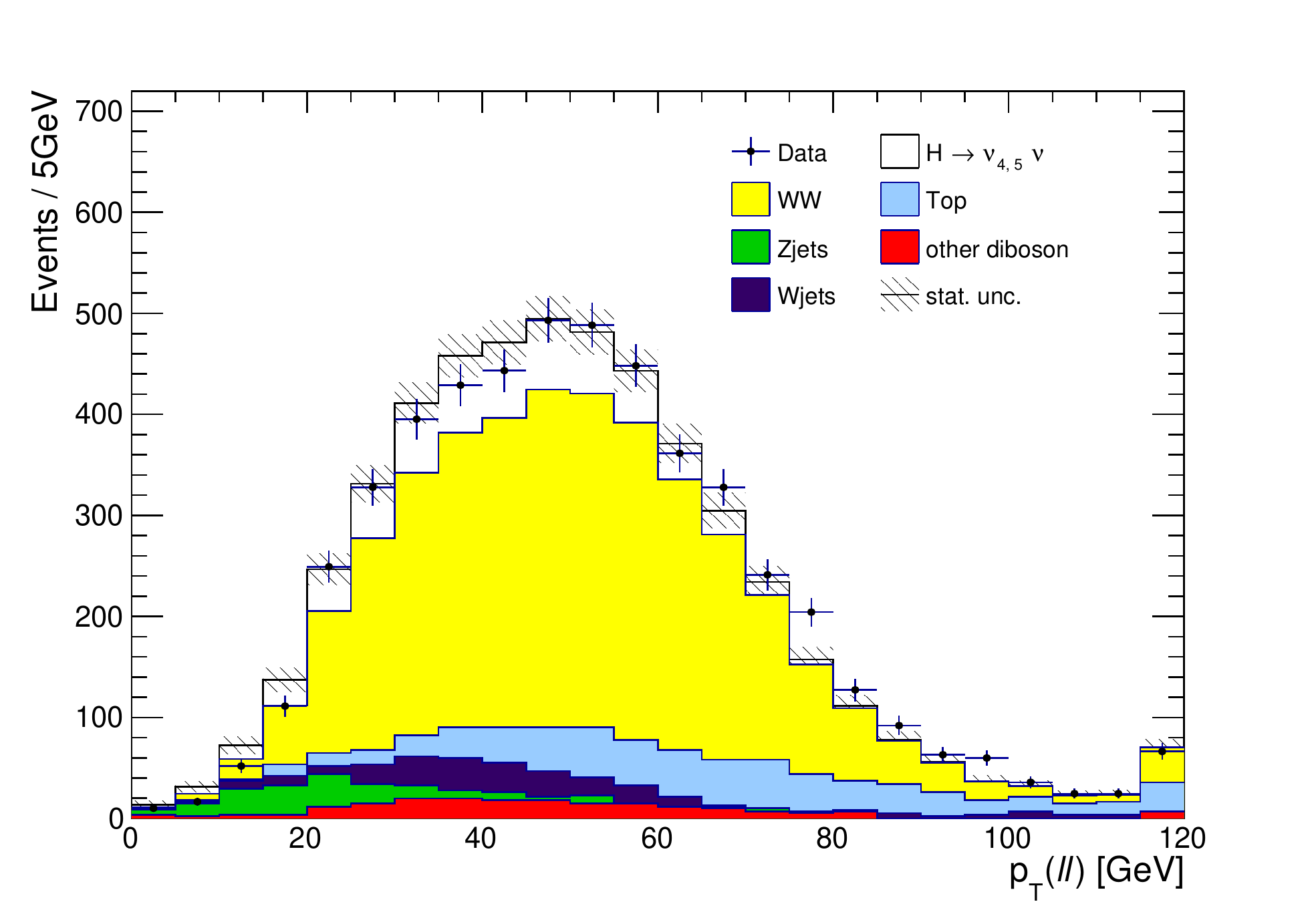}
\includegraphics[width=.49\linewidth]{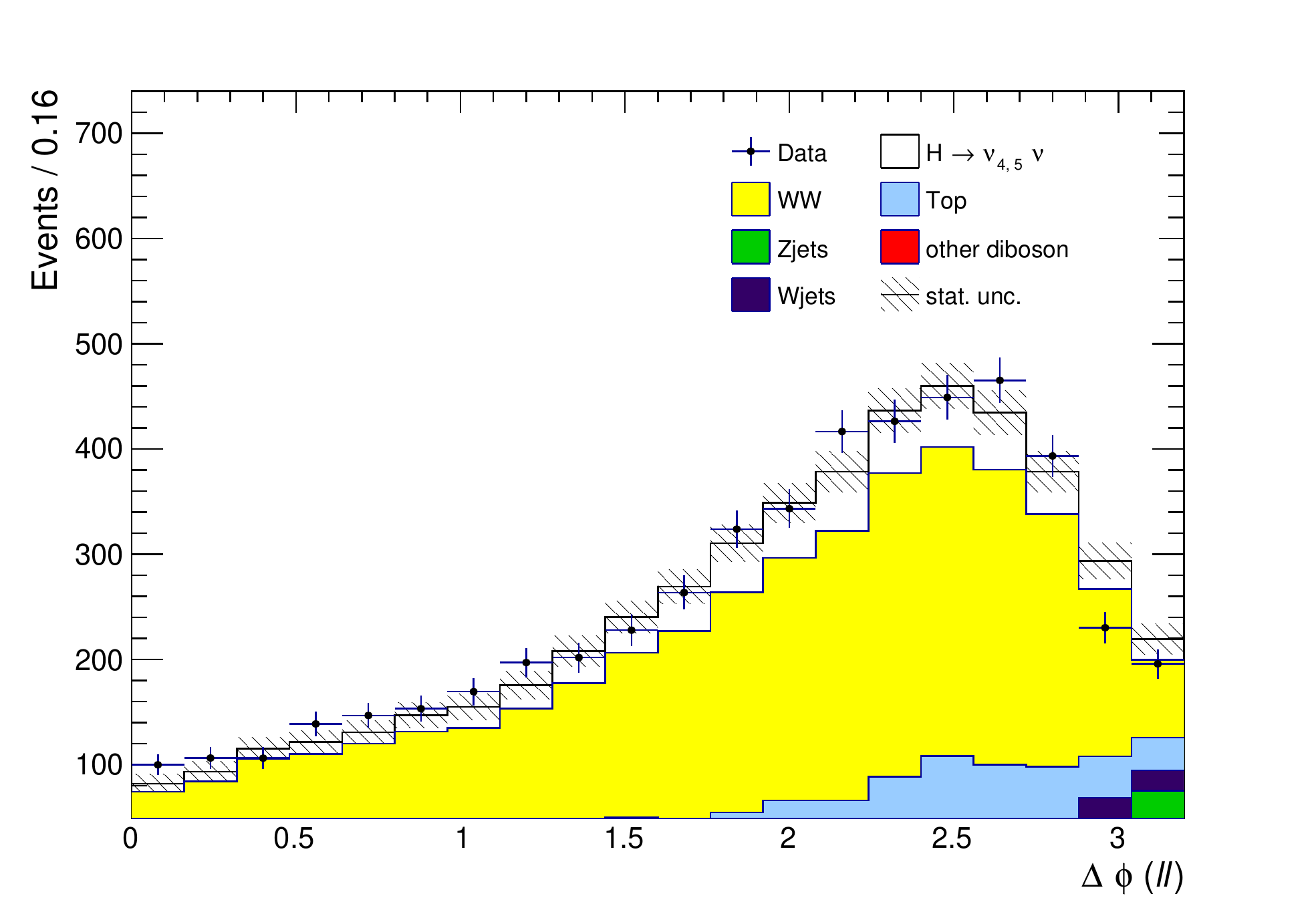}
\includegraphics[width=.49\linewidth]{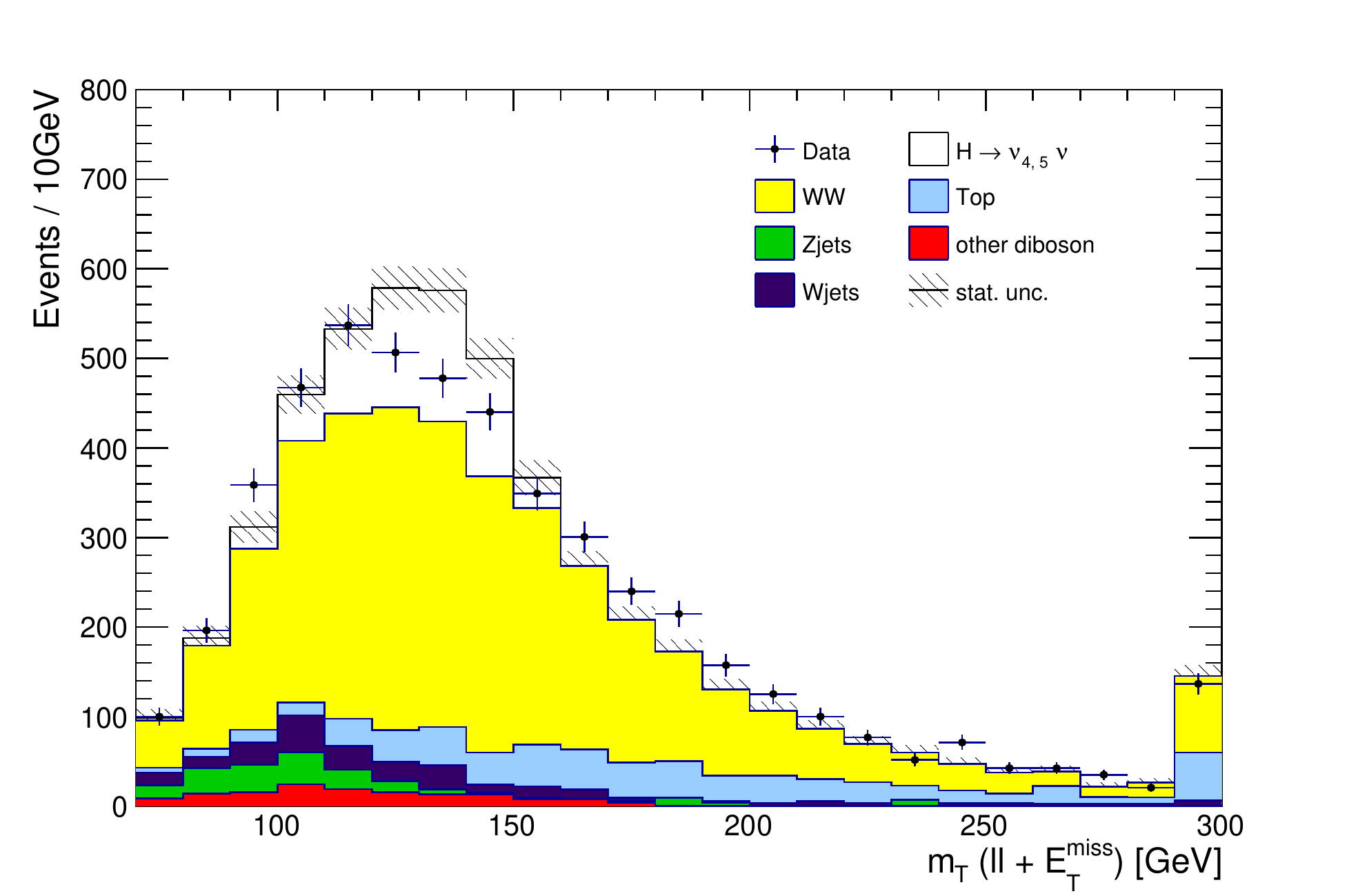}
\includegraphics[width=.49\linewidth]{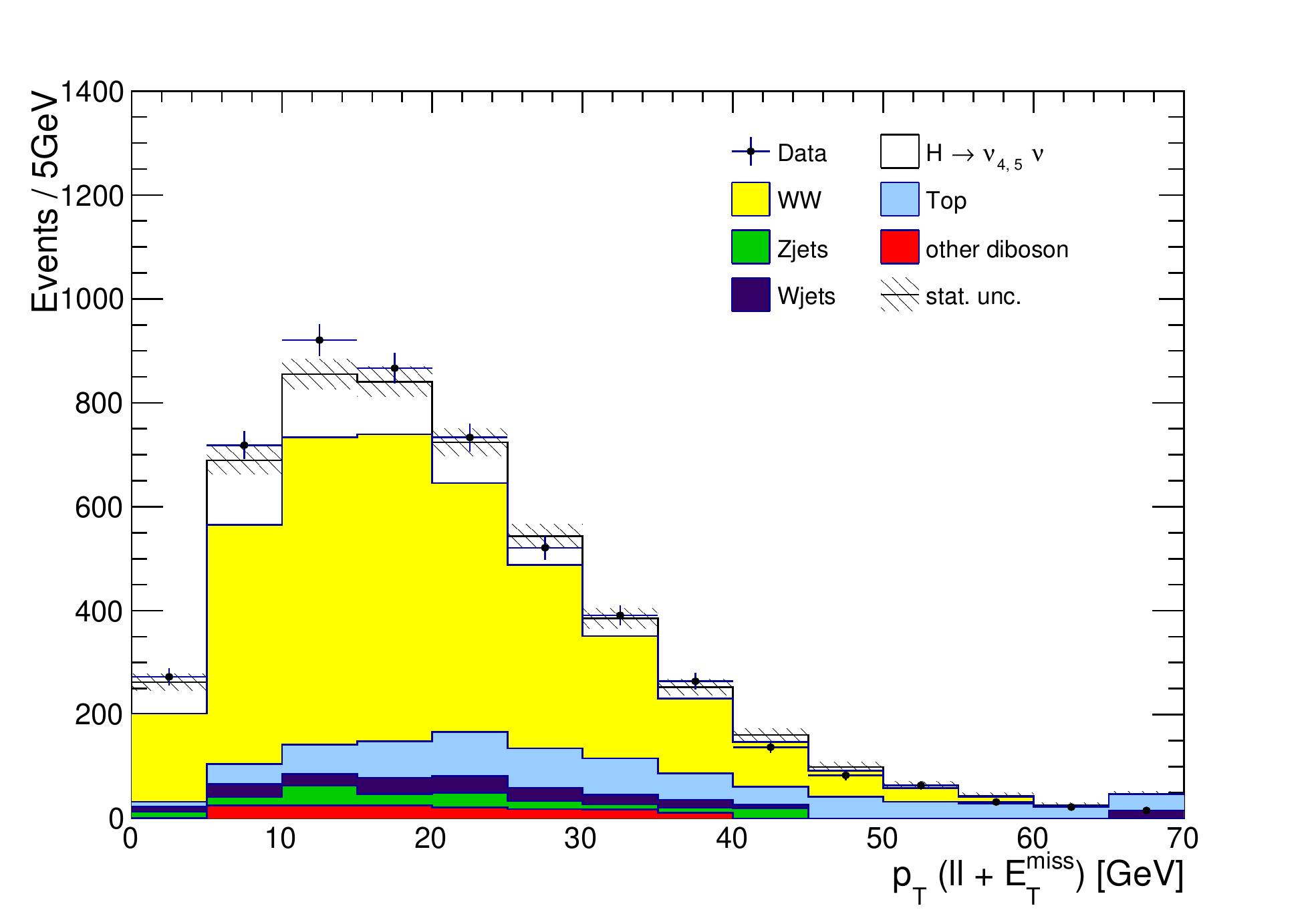}
\caption{Kinematic distributions for $pp \to H \to \nu_4 \nu_\mu,\nu_5 \nu_e \to e \mu \nu_e \nu_\mu$ corresponding to our reference  parameters $m_H = 155 \; {\rm GeV}$, $m_{\nu_{4,5}} = 135 \; {\rm GeV}$, ${\rm BR} (H\to  W\ell\nu_\ell) = 0.16$. The effective cross section is about 90\% of the required contribution. The acceptance we find is $A_{\rm NP}$ = 26.6\%. \label{fig:emu}}
\end{center}
\end{figure}
\begin{figure}
\begin{center}
\includegraphics[width=.49\linewidth]{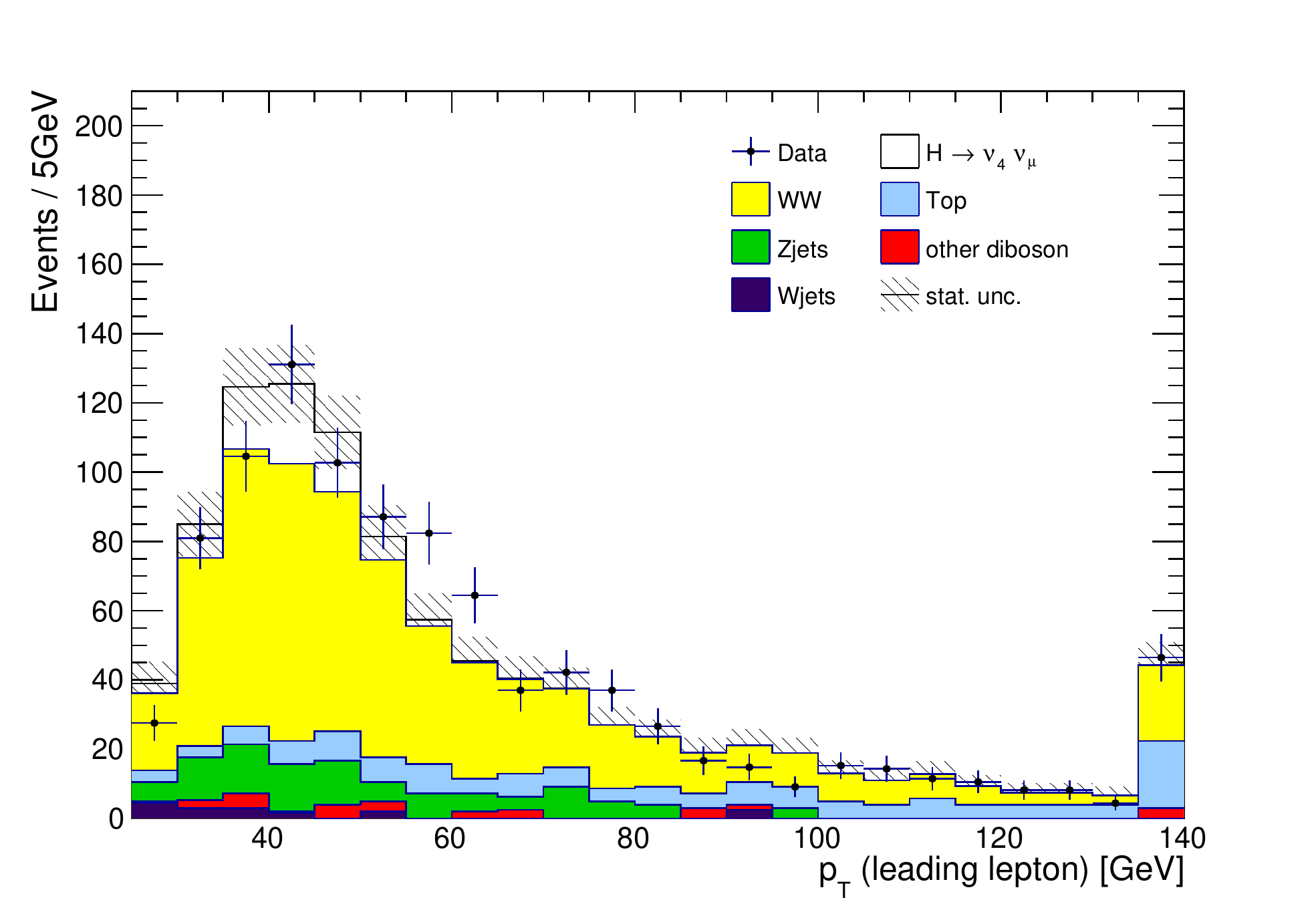}
\includegraphics[width=.49\linewidth]{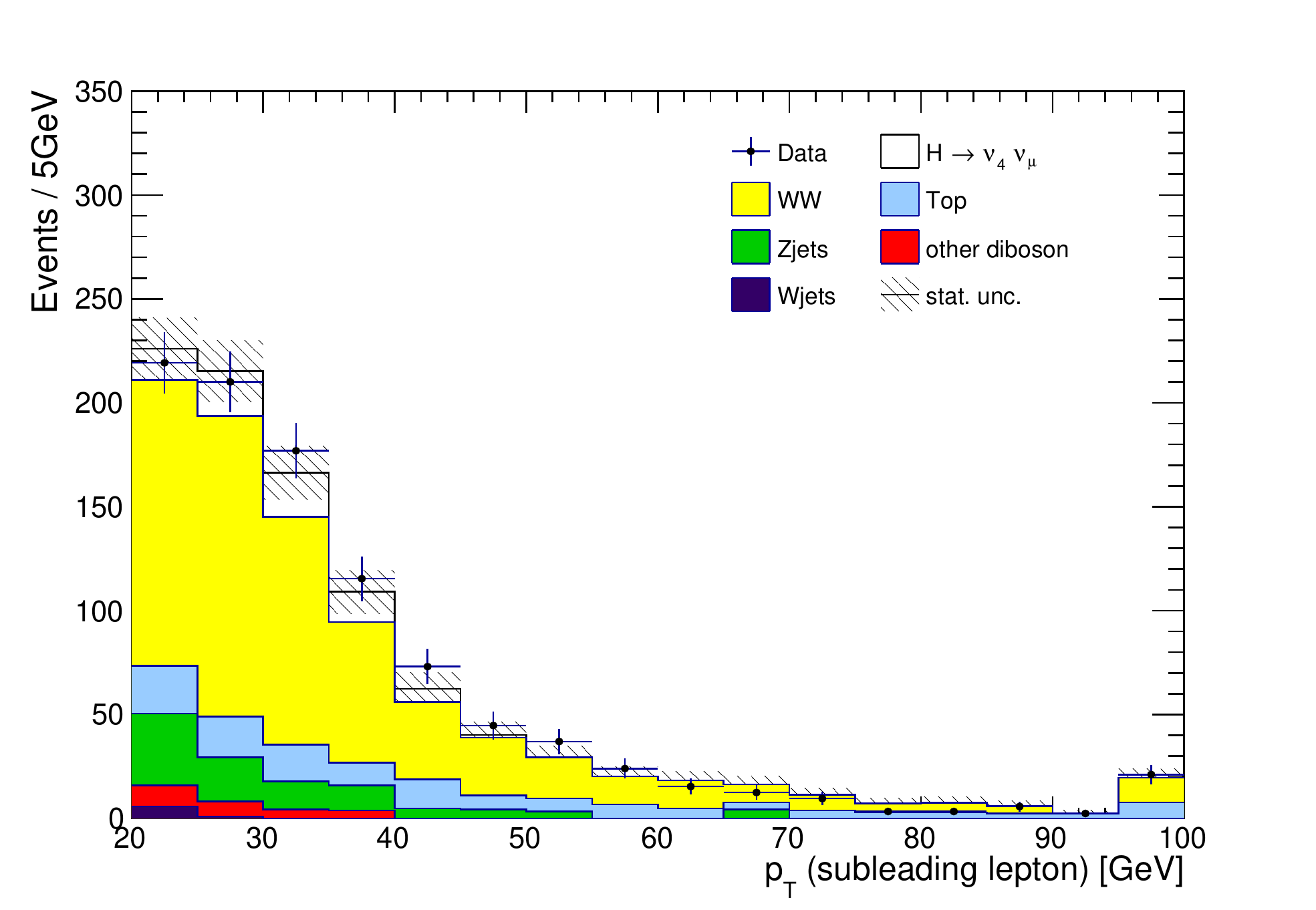}
\includegraphics[width=.49\linewidth]{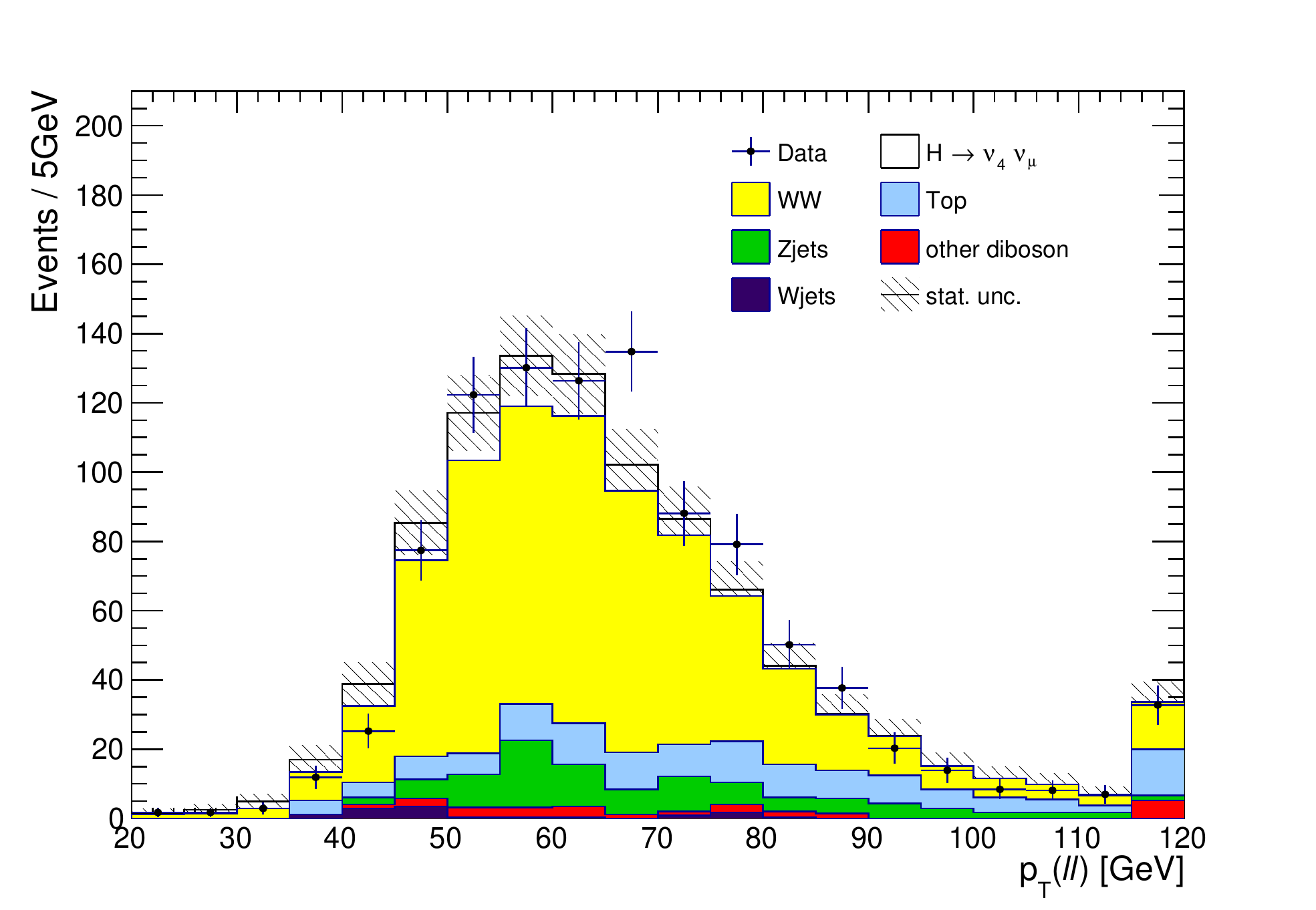}
\includegraphics[width=.49\linewidth]{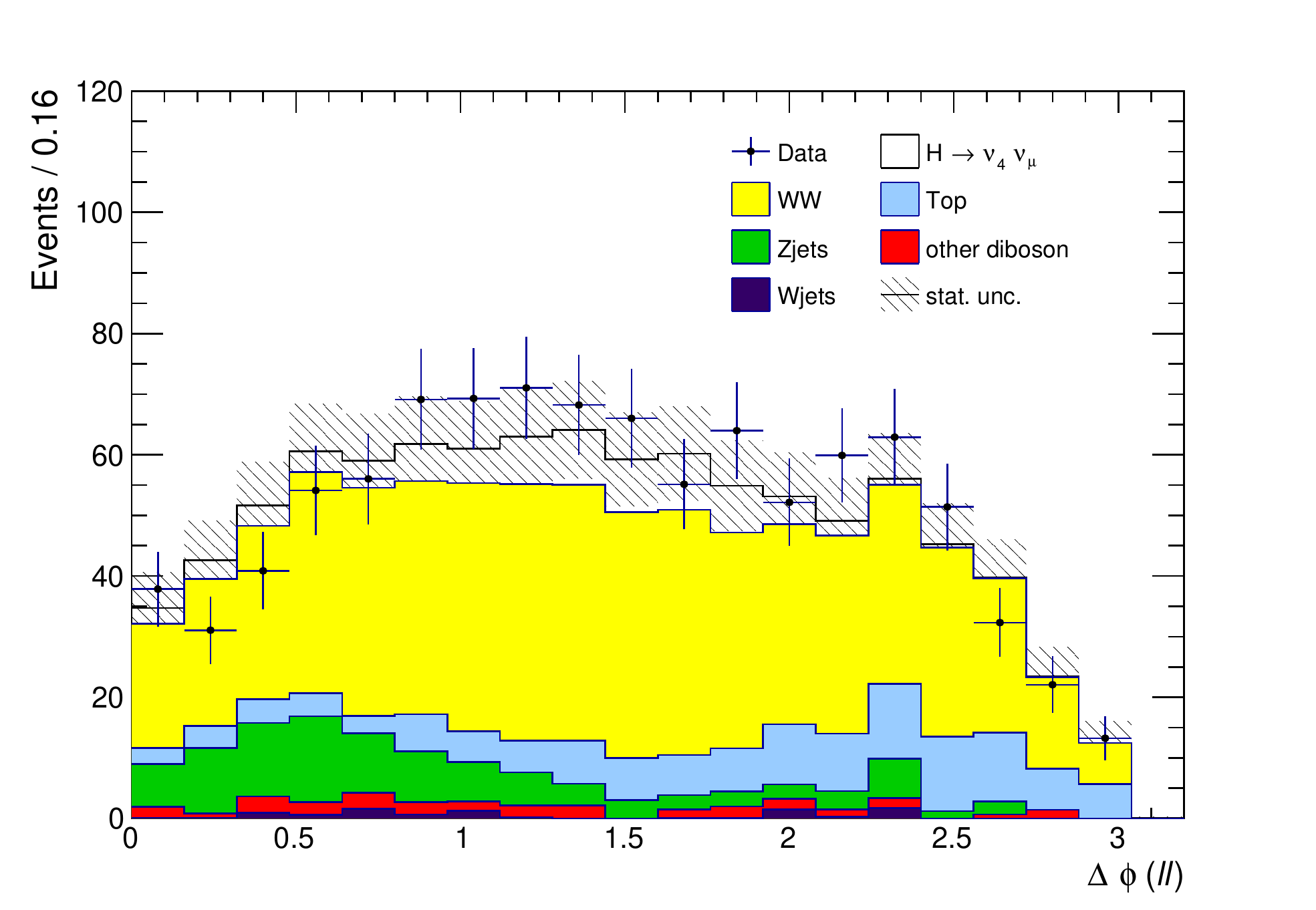}
\includegraphics[width=.49\linewidth]{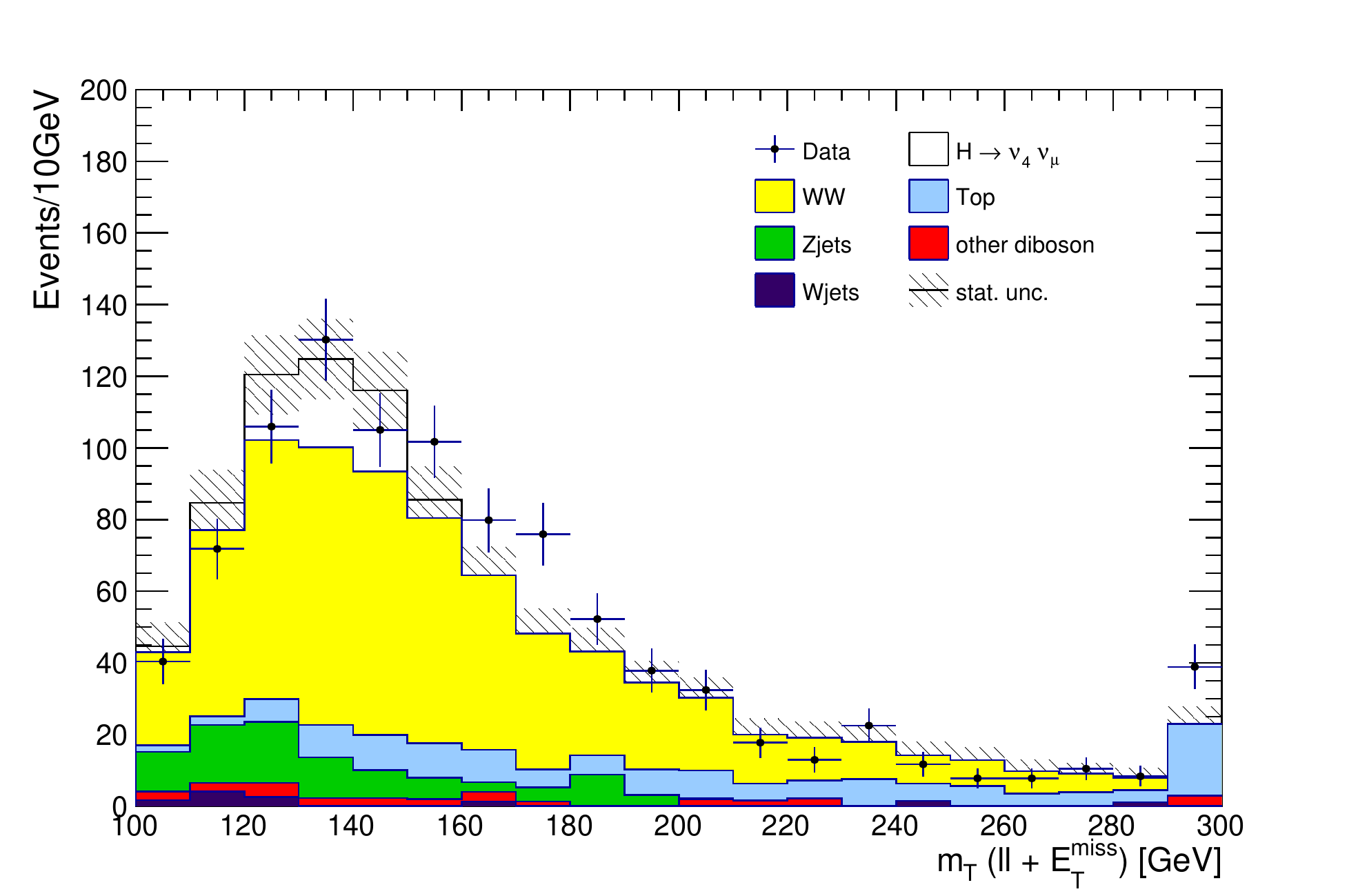}
\includegraphics[width=.49\linewidth]{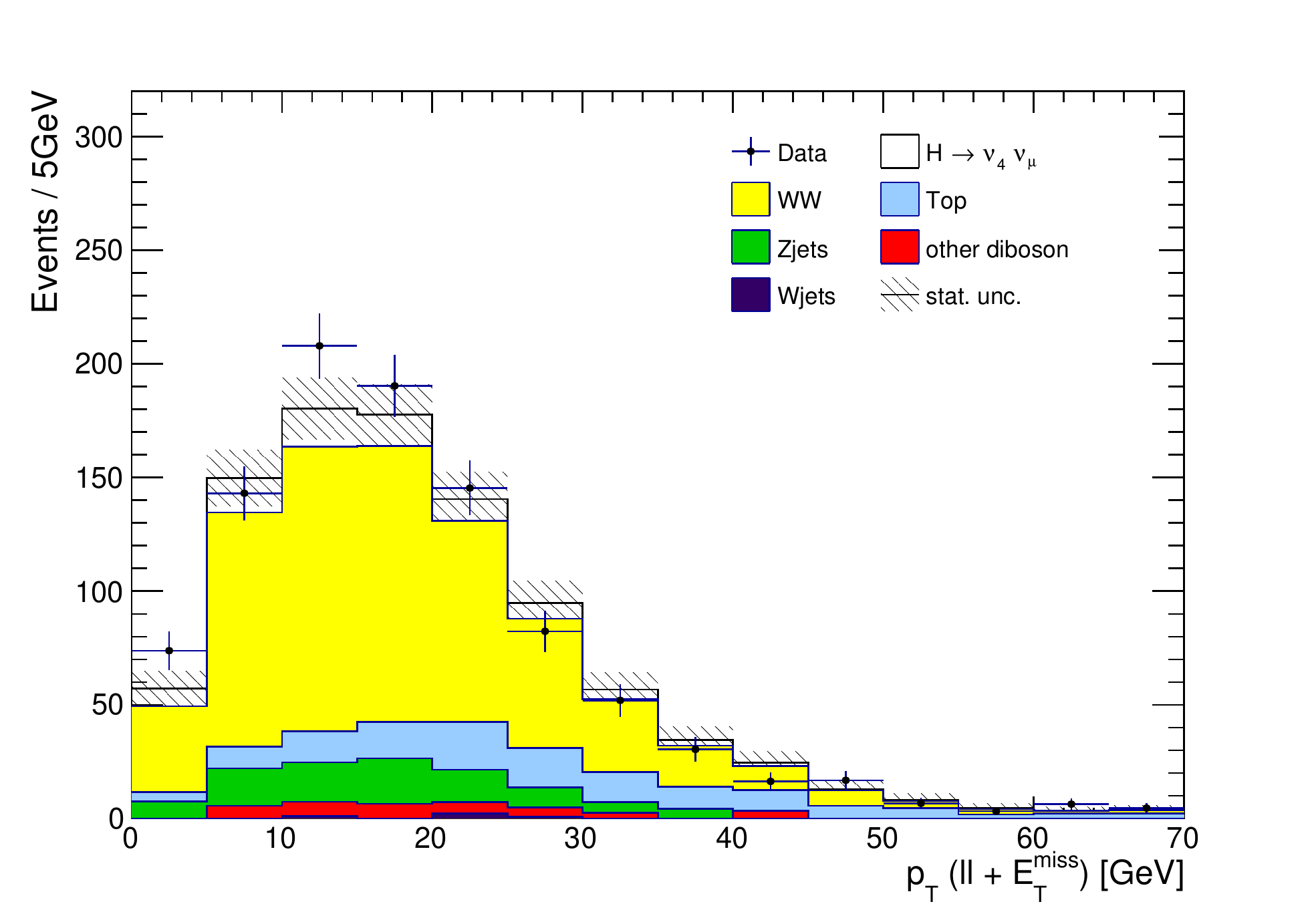}
\caption{Kinematic distributions for $pp \to H \to \nu_{4} \nu_{\mu} \to \mu \mu \nu_\mu \nu_\mu$ corresponding to our reference  parameters $m_H = 155 \; {\rm GeV}$, $m_{\nu_{4}} = 135 \; {\rm GeV}$, ${\rm BR} (H\to  W\ell\nu_\ell) = 0.08$. The effective cross section is about 75\% of the required contribution. The acceptance we find is $A_{\rm NP}$ = 7.2\%. The corresponding distributions for the $ee$ channel are scaled down by a factor $C_{WW}^{ee}/C_{WW}^{\mu\mu} = 0.62$ (see table~5 of ref.~\cite{atlasww}). \label{fig:mumu}}
\end{center}
\end{figure}
\begin{figure}
\begin{center}
\includegraphics[width=.49\linewidth]{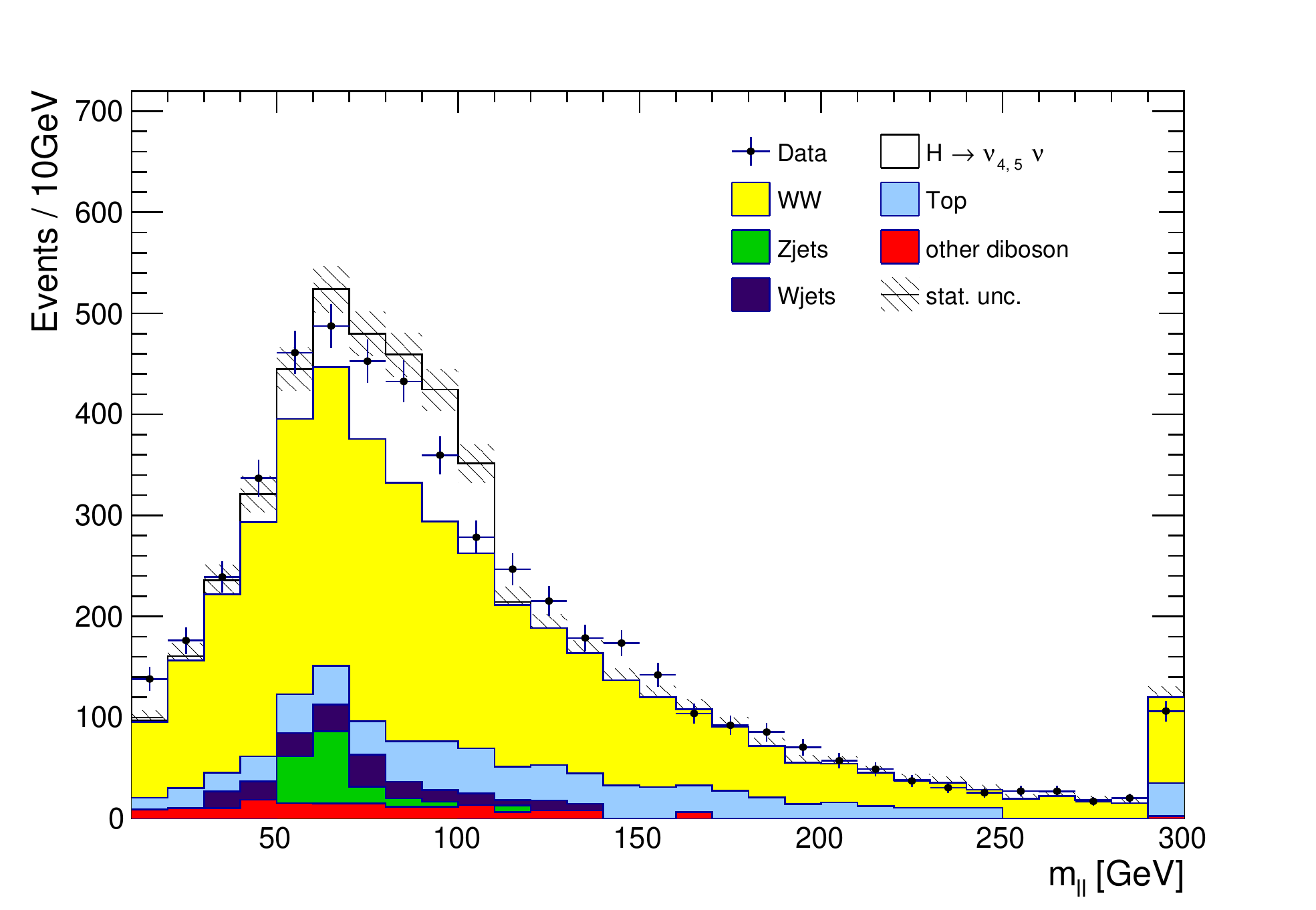}
\includegraphics[width=.49\linewidth]{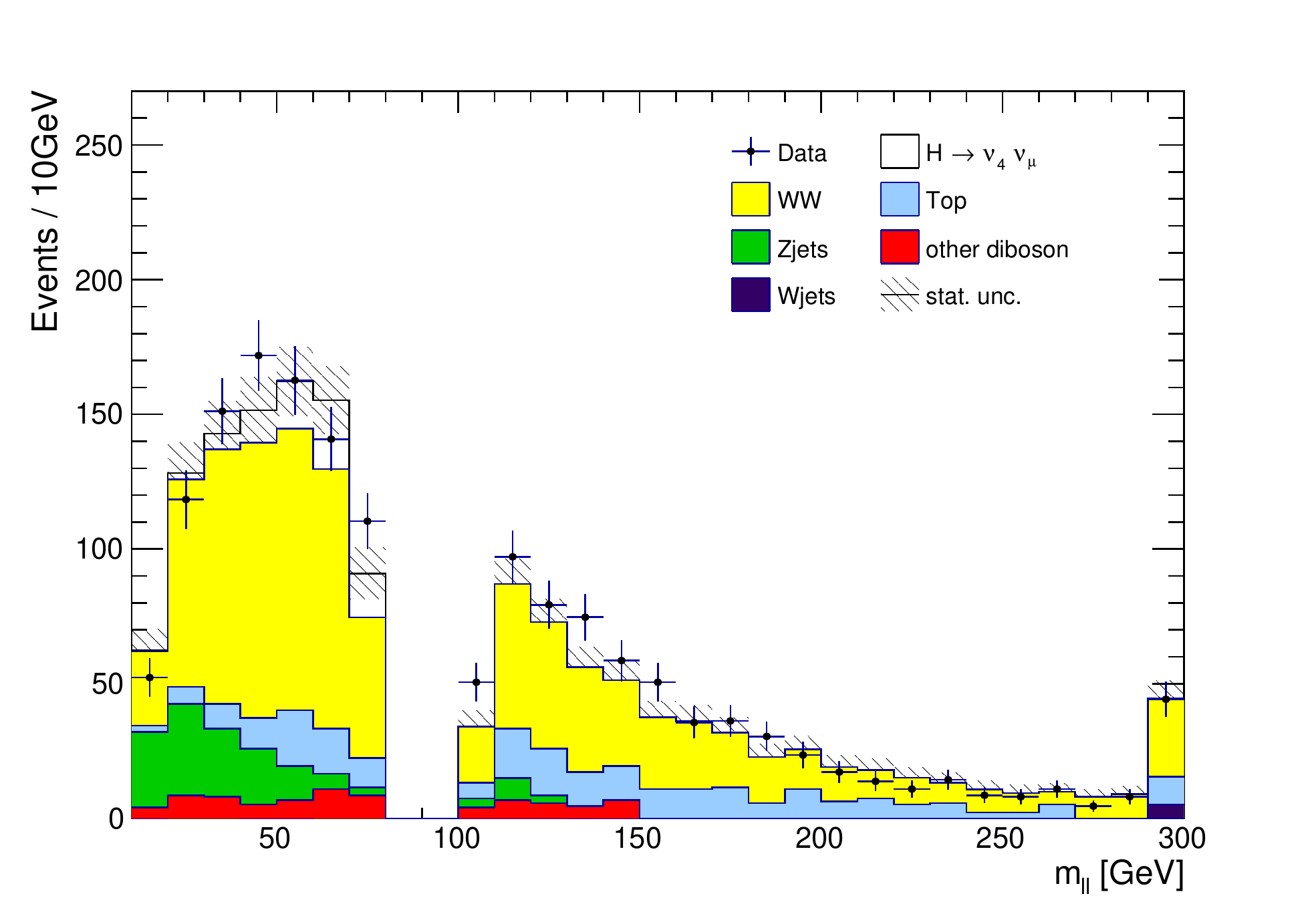}
\caption{Distribution of the dilepton invariant mass $m_{\ell \ell}$ for the $e\mu$ and $\mu\mu$ channels. See Figs.~\ref{fig:emu} and \ref{fig:mumu} for further details. \label{fig:mll}}
\end{center}
\end{figure}

In order to compare with the experimental results presented in ref.~\cite{atlasww}, we convert our (differential) fiducial cross sections into number of events:
\begin{align}
N_{\rm NP} = \sigma_{\rm NP}^{\rm fid} \; {\cal L} \; C_{WW}
\end{align}
where ${\cal L} = 20.3 \; {\rm fb}^{-1}$ is the ATLAS integrated luminosity and $C_{WW}$ is a correction factor that takes into account detector effects (this factor is explicitly given in ref.~\cite{atlasww} for the $e\mu$, $ee$ and $\mu\mu$ modes, separately). The main reason for using the observed number of events is that, in this way, we can easily assess statistical uncertainties on our signal in reference to the observed excess. 

We select a representative point with $m_H = 155 \; {\rm GeV}$, $m_{\nu_{4,5}} = 135 \; {\rm GeV}$ and ${\rm BR}(H \to  W \ell\nu_\ell ) = 0.16$ for the $e\mu$ mode and 0.08 for the $ee$ and $\mu\mu$ ones. The distributions that we obtain are presented in figures~\ref{fig:emu}--\ref{fig:mll}. The product of branching ratios has been chosen to obtain an effective cross section of about 11.5 pb ( 90\% of the $e\mu$ excess), corresponding to about 7.5 pb for $\mu\mu$ case (75\% of the $\mu\mu$ excess). Note that in figures~\ref{fig:emu}--\ref{fig:mll} all background estimates are taken directly from ref.~\cite{atlasww} while the new physics signal is simulated with MadGraph5 interfaced with Pythia (with detector effects taken into account via the factors $C_{WW}$). As a consistency check we simulated $pp\to WW$ events using the same framework we use for the signal and found that all kinematic distributions agree fairly well with those presented in ref.~\cite{atlasww} (obtained using POWHEG interfaced with Pythia).

Possible additional freedom from considering different masses and branching ratios related to two new leptons  would allow to adjust contributions to the $ee$ and $\mu\mu$ modes separately and to tweak various kinematic distributions. At present, this is however not necessary since  ATLAS finds similar effects in both modes - see, for instance, table~\ref{table:atlasxsec}, and all the kinematic distributions are fit nicely with the assumptions of universality in masses and couplings of $\nu_4$ and $\nu_5$.

Direct inspection of these figures shows that all the kinematic distributions we consider agree perfectly with the ATLAS observations with the exception of few bins in the transverse mass, $m_T (\ell\ell + E_T^{\rm miss})$, distribution. The choice of the reference point has been driven by this variable for the $e\mu$ channel (the $ee$ and $\mu\mu$ ones showing much less significant excesses). In the new physics process we consider, contributions to $m_T$ are bounded from above by the mass of the heavy Higgs. ATLAS data show a large excess in the $m_T \in [90,150]\; {\rm GeV}$ range and a moderate/small one at larger $m_T$. Choosing $m_H = 155 \; {\rm GeV}$ guarantees that we are able to explain the bulk of the excess. The mass of the heavy neutral lepton has a subleading effect on $m_T$ but controls entirely the $p_T (\ell \ell)$ distribution. We chose $m_{\nu_{4,5}} = 135 \; {\rm GeV}$ but we would like to stress that we obtain similar results with different choices of masses. 

In most of the preferred region in figure~\ref{fig:MhMnu} all the kinematic distributions except $m_T$ typically agree well with observations for proper choices of ${\rm BR} (H\to W\ell\nu_\ell)$. For Higgs masses heavier than 160 GeV, $m_T$ still agrees fairly well with the observed excess but the distribution broadens and moves to somewhat higher $m_T$ bins. 

Finally we note that if at least one of the heavy neutral leptons is lighter than the SM Higgs (125 GeV) then it is reasonable to expect a non-zero BR($h_{\rm SM} \to \nu_{4,5} \nu$). In this case both $H$ and $h_{\rm SM}$ contribute to the excess through the same decay chains. The $m_T$ distribution of the SM Higgs contribution is localized at low energies ($m_T < 125$ GeV) while the other distributions are not significantly affected. While we checked that the $h_{\rm SM}$ contribution alone is unable to fully account for the $WW$ excess, the combination of $H$ and $h_{\rm SM}$ should clearly work better than $H$ alone. 

\section{Conclusions}
\label{sec:conclusions}

In this paper we have investigated contributions to $pp\to W^+W^- \to \ell \nu_\ell \ell^\prime \nu_{\ell^\prime}$ in models with a new Higgs boson, $H$, and a neutral lepton, $\nu_{4}$, with couplings $H-\nu_{4}-\nu_{\mu}$ and $W-\nu_{4}-\mu$ through the process $pp \to H \to \nu_4 \nu_\mu \to W \mu \nu_\mu  \to \ell\nu_\ell \mu \nu_\mu$. These contributions can be very large and thus the cross sections measured at ATLAS~\cite{atlasww} and CMS~\cite{CMS:2015uda} offer powerful constraints. This scenario is able to generate large contributions because of a Higgs production cross section of order 10 pb coupled with having only one $W$ in the final state (and hence avoiding the double $W$ leptonic branching ratio suppression).

Contrary to naive expectations, we find that in a wide range of masses and branching ratios the present sensitivity of $pp\to WW$ measurements offers stronger constraints than direct searches for heavy Higgses in the $WW$ and $\gamma \gamma$ channels. $H\to W W$ constraints are weakened by the strong suppression of the ratio of acceptances $A_{\rm NP}^{\cal H} / A_{\rm NP}$ (see eq.~(\ref{eq:WWbounds_simp})). $H\to\gamma\gamma$ searches are diluted by the requirement that the heavy Higgs has no direct couplings to the SM gauge bosons. Our main results are summarized in figure~\ref{fig:MhMnu}.

In addition we selected a representative point in the parameter space for which the NP contribution to the fiducial cross section matches roughly the excess observed by ATLAS. We studied several kinematic variables and found that all observed distributions can be easily accommodated in our scenario. 

Future experimental updates of $pp \to WW$ measurements will be crucial to test the class of models studied in this paper. Furthermore, in the region of the parameter space in which we have significant contributions to $WW$, our scenario necessarily predicts deviations in searches for heavy Higgses into $\ell \nu_\ell \ell^\prime \nu_{\ell^\prime}$ and $\gamma\gamma$ final states that might be observable in the near future. 

In addition to the results presented in this paper, we also investigated several alternatives. One possibility is to generate simultaneous contributions in the $e\mu$, $ee$ and $\mu\mu$ channels by coupling only the tau (rather than $e$ or $\mu$) neutrino to the new lepton. The decay $H\to \nu_4 \nu_\tau \to W\tau \nu_\tau$ with a leptonically decaying tau produces all three final states. In this case, the branching ratios for leptonic $\tau$ decays reduce the overall cross section by a factor $\sim 5$. Furthermore, the extra neutrinos produced in these decays result in additional missing energy, thus lowering the $p_T$ of the charged leptons. This makes it hard for these events to pass the ATLAS and CMS selection cuts and we found that the resulting rates are low. We also considered direct Drell-Yan lepton production (e.g. $pp \to W \to \nu_4\mu$) and found that it also yields small cross sections.

The new neutral leptons can originate from extensions of the SM by vectorlike leptons, both SU(2) doublets and neutral singlets in a two Higgs doublet model framework. In any specific model there are additional constraints on masses and couplings of the new leptons. These include constraints from electroweak precision data and from pair production of new leptons from searches for anomalous production of multi lepton events, discussed in ref.~\cite{Dermisek:2014qca}. We will discuss an explicit scenario along these lines in a forthcoming publication~\cite{Dermisek:2015oja}.

\acknowledgments 
E.L. thanks Giulia Zanderighi and Stefano Pozzorini for useful discussions and clarifications. This work is supported in part by the Department of Energy under grant number {DE}-SC0010120.

\appendix

\section{Detailed description of $H\to W W$ constraints}
\label{app}
In refs.~\cite{CMS:bxa, Chatrchyan:2013iaa} CMS considered a large number of different cuts each optimized to be sensitive to a SM--like heavy Higgs of a given mass. The signal we consider here ($H \to \nu_{4} \nu_\mu \to \mu \nu_{\mu} \ell\nu_\ell$) is topologically different from the SM Higgs decay ($H\to WW\to \mu \nu_{\mu} \ell\nu_\ell$) and cuts optimized for SM Higgs hypotheses are not in general optimal for our process. For every point in the new physics parameter space, i.e. $m_H$, $m_{\nu_{4}}$, ${\rm BR}(H\to W\ell\nu_\ell)$, we consider the constraints implied by each CMS analysis and take the strongest bound we obtain. Each CMS analysis, that we indicate by ${\cal H}$, has been optimized for a given SM Higgs mass hypothesis $\hat m_H$. 

The number of surviving new physics events for our signal ($N_{\rm NP}^{\cal H}$) and for a SM like heavy Higgs ($N_{\rm SM}^{\cal H}$) that we expect are given by
\begin{align}
N_{\rm NP}^{\cal H} (m_H, m_{\nu_{4}}, {\rm BR}) &= A_{\rm NP}^{\cal H} (m_H,m_{\nu_{4}}) \; \sigma_{\rm NP} (m_H,m_{\nu_{4}},{\rm BR}) \; C^{\cal H} \; \mathcal{L} \; , \label{eq:NNPH}\\
N_{\rm SM}^{\cal H} (\hat m_H) & = A_{\rm SM}^{\cal H} (\hat m_H) \; \sigma_{\rm SM} (\hat m_H) \; C^{\cal H} \; \mathcal{L} \; ,
\label{eq:NSMH}
\end{align}
where $\sigma_{\rm NP}$ and $\sigma_{\rm SM}$ are the total cross sections (including branching ratios) as in eqs.~(\ref{sigma_fiducial_SM}) and (\ref{sigma_fiducial_NP}), $A_{\rm NP}^{\cal H}$ and $A_{\rm SM}^{\cal H}$ are the corresponding acceptances, ${\cal L}$ is the total integrated luminosity and $C^{\cal H}$ encapsulates detector efficiencies. We consider the mass $\hat m_H$ in eq.~(\ref{eq:NSMH}) because all the quantities that thus appear are explicitly given in refs.~\cite{CMS:bxa, Chatrchyan:2013iaa} and we are therefore able to extract the product $C^{\cal H} {\cal L}$. Combining eqs.~(\ref{eq:NNPH}) and (\ref{eq:NSMH}) we obtain
\begin{align}
N_{\rm NP}^{\cal H} (m_H, m_{\nu_{4}}, {\rm BR}) &= 
\frac{A_{\rm NP}^{\cal H} (m_H,m_{\nu_{4}})}{A_{\rm SM}^{\cal H} (\hat m_H) } \; 
\frac{\sigma_{\rm NP} (m_H,m_{\nu_{4}},{\rm BR})}{\sigma_{\rm SM} (\hat m_H)} \;
N_{\rm SM}^{\cal H} (\hat m_H) \; .
\label{eq:NNPHfinal}
\end{align}
The next step is to extract the upper limit at 95\% C.L. that the observed number of events ($N_{\rm exp}^{\cal H}$) and the complete (including the 125 GeV Higgs) SM background ($N_{\rm bkgd}^{\cal H}$) imply. We follow a standard ${\rm CL}_s$ method (see Appendix D in ref.~\cite{Dermisek:2013cxa} for a detailed description of the technique) at obtain the upper bounds $N_{\rm NP}^{\cal H} < \ell_{95}^{\cal H}$ (we add a 30\% systematic uncertainty to our signal in addition to the standard gaussian statistical error). Therefore, the upper limit on the new physics cross section is
\begin{align}
\sigma_{\rm NP} (m_H,m_{\nu_{4}},{\rm BR}) &< \min_\mathcal{H} \left[
\frac{1}{A_{\rm NP}^{\cal H} (m_H,m_{\nu_{4}})} 
\frac{A_{\rm SM}^{\cal H} (\hat m_H)\; \sigma_{\rm SM} (\hat m_H) \; \ell_{95}^{\cal H}}{N_{\rm SM}^{\cal H} (\hat m_H)}
\right] \\
&\equiv \min_\mathcal{H} \left[ \frac{\beta^{\cal H}_{95}}{A_{\rm NP}^{\cal H} (m_H,m_{\nu_{4}})}
\right] \; .
\end{align}
The quantities $\beta_{95}^{\cal H}$ are independent of new physics parameters, can be extracted entirely from refs.~\cite{CMS:bxa, Chatrchyan:2013iaa} and are listed in table~\ref{tab:beta}. The implied upper limit on the fiducial cross section is
\begin{align}
\sigma^{\rm fid}_{\rm NP} (m_H,m_{\nu_{4}},{\rm BR}) &< A_{\rm NP}(m_H,m_{\nu_{4}}) \; 
\min_\mathcal{H} \left[ \frac{\beta^{\cal H}_{95}}{A_{\rm NP}^{\cal H} (m_H,m_{\nu_{4}})} \right] \; .
\label{eq:WWbounds}
\end{align}


\begin{thebibliography}{99}

\bibitem{atlasww} 
  The ATLAS collaboration,
  ATLAS-CONF-2014-033, ATLAS-COM-CONF-2014-045.

\bibitem{Nason:2004rx} 
  P.~Nason,
  JHEP {\bf 0411}, 040 (2004)
  [hep-ph/0409146].

\bibitem{Frixione:2007vw} 
  S.~Frixione, P.~Nason and C.~Oleari,
  JHEP {\bf 0711}, 070 (2007)
  [arXiv:0709.2092 [hep-ph]].

\bibitem{Alioli:2010xd} 
  S.~Alioli, P.~Nason, C.~Oleari and E.~Re,
  JHEP {\bf 1006}, 043 (2010)
  [arXiv:1002.2581 [hep-ph]].

\bibitem{Melia:2011tj} 
  T.~Melia, P.~Nason, R.~Rontsch and G.~Zanderighi,
  JHEP {\bf 1111}, 078 (2011)
  [arXiv:1107.5051 [hep-ph]].

\bibitem{Nason:2013ydw} 
  P.~Nason and G.~Zanderighi,
  Eur.\ Phys.\ J.\ C {\bf 74}, no. 1, 2702 (2014)
  [arXiv:1311.1365 [hep-ph]].

\bibitem{Campbell:2011bn} 
  J.~M.~Campbell, R.~K.~Ellis and C.~Williams,
  JHEP {\bf 1107}, 018 (2011)
  [arXiv:1105.0020 [hep-ph]].

\bibitem{cmswwold}
  S.~Chatrchyan et al.  [CMS Collaboration],
  Phys.\ Lett.\ B {\bf 721}, 190 (2013)
  [arXiv:1301.4698 [hep-ex]].


\bibitem{Heinemeyer:2013tqa} 
  S.~Heinemeyer {\it et al.}  [LHC Higgs Cross Section Working Group Collaboration],
  arXiv:1307.1347 [hep-ph].

\bibitem{Curtin:2012nn} 
  D.~Curtin, P.~Jaiswal and P.~Meade,
  Phys.\ Rev.\ D {\bf 87}, no. 3, 031701 (2013)
  [arXiv:1206.6888 [hep-ph]].

\bibitem{Jaiswal:2013xra} 
  P.~Jaiswal, K.~Kopp and T.~Okui,
  Phys.\ Rev.\ D {\bf 87}, no. 11, 115017 (2013)
  [arXiv:1303.1181 [hep-ph]].

\bibitem{Rolbiecki:2013fia} 
  K.~Rolbiecki and K.~Sakurai,
  JHEP {\bf 1309}, 004 (2013)
  [arXiv:1303.5696 [hep-ph]].

\bibitem{Curtin:2013gta} 
  D.~Curtin, P.~Jaiswal, P.~Meade and P.~J.~Tien,
  JHEP {\bf 1308}, 068 (2013)
  [arXiv:1304.7011 [hep-ph]].

\bibitem{Curtin:2014zua} 
  D.~Curtin, P.~Meade and P.~J.~Tien,
  Phys.\ Rev.\ D {\bf 90}, no. 11, 115012 (2014)
  [arXiv:1406.0848 [hep-ph]].

\bibitem{Kim:2014eva} 
  J.~S.~Kim, K.~Rolbiecki, K.~Sakurai and J.~Tattersall,
  JHEP {\bf 1412}, 010 (2014)
  [arXiv:1406.0858 [hep-ph]].

\bibitem{CMS:2015uda} 
  CMS Collaboration [CMS Collaboration],
  CMS-PAS-SMP-14-016.

\bibitem{Meade:2014fca} 
  P.~Meade, H.~Ramani and M.~Zeng,
  Phys.\ Rev.\ D {\bf 90}, no. 11, 114006 (2014)
  [arXiv:1407.4481 [hep-ph]].
  
\bibitem{Jaiswal:2014yba} 
  P.~Jaiswal and T.~Okui,
  Phys.\ Rev.\ D {\bf 90}, no. 7, 073009 (2014)
  [arXiv:1407.4537 [hep-ph]].

\bibitem{Gehrmann:2014fva} 
  T.~Gehrmann, M.~Grazzini, S.~Kallweit, P.~Maierhöfer, A.~von Manteuffel, S.~Pozzorini, D.~Rathlev and L.~Tancredi,
  Phys.\ Rev.\ Lett.\  {\bf 113}, no. 21, 212001 (2014)
  [arXiv:1408.5243 [hep-ph]].

\bibitem{Monni:2014zra} 
  P.~F.~Monni and G.~Zanderighi,
  arXiv:1410.4745 [hep-ph].
  
\bibitem{Dermisek:2013gta} 
  R.~Dermisek and A.~Raval,
  Phys.\ Rev.\ D {\bf 88}, 013017 (2013)
  [arXiv:1305.3522 [hep-ph]].
 
\bibitem{Dermisek:2014cia} 
  R.~Dermisek, A.~Raval and S.~Shin,
  Phys.\ Rev.\ D {\bf 90}, no. 3, 034023 (2014)
  [arXiv:1406.7018 [hep-ph]].
  
\bibitem{Dermisek:2015oja} 
  R.~Dermisek, E.~Lunghi and S.~Shin,
  arXiv:1509.04292 [hep-ph].
  
\bibitem{BhupalDev:2012zg} 
  P.~S.~Bhupal Dev, R.~Franceschini and R.~N.~Mohapatra,
  Phys.\ Rev.\ D {\bf 86}, 093010 (2012)
  [arXiv:1207.2756 [hep-ph]].

\bibitem{CMS:bxa} 
  CMS Collaboration,
  CMS-PAS-HIG-13-003.

\bibitem{Chatrchyan:2013iaa}   
  S.~Chatrchyan et al.  [CMS Collaboration],
  JHEP {\bf 1401}, 096 (2014)
  [arXiv:1312.1129 [hep-ex]].

\bibitem{CMS:2014onr} 
  CMS Collaboration,
  CMS-PAS-HIG-14-006.

\bibitem{Dermisek:2014qca} 
  R.~Dermisek, J.~P.~Hall, E.~Lunghi and S.~Shin,
  JHEP {\bf 1412}, 013 (2014)
  [arXiv:1408.3123 [hep-ph]].

\bibitem{Alwall:2014hca} 
  J.~Alwall, R.~Frederix, S.~Frixione, V.~Hirschi, F.~Maltoni, O.~Mattelaer, H.-S.~Shao and T.~Stelzer et al.,
  JHEP {\bf 1407}, 079 (2014)
  [arXiv:1405.0301 [hep-ph]].

\bibitem{Sjostrand:2006za} 
  T.~Sjostrand, S.~Mrenna and P.~Z.~Skands,
  JHEP {\bf 0605}, 026 (2006)
  [hep-ph/0603175].

\bibitem{deFavereau:2013fsa} 
  J.~de Favereau {\it et al.}  [DELPHES 3 Collaboration],
  JHEP {\bf 1402}, 057 (2014)
  [arXiv:1307.6346 [hep-ex]].

\bibitem{Cacciari:2005hq} 
  M.~Cacciari and G.~P.~Salam,
  Phys.\ Lett.\ B {\bf 641}, 57 (2006)
  [hep-ph/0512210].

\bibitem{Cacciari:2011ma} 
  M.~Cacciari, G.~P.~Salam and G.~Soyez,
  Eur.\ Phys.\ J.\ C {\bf 72}, 1896 (2012)
  [arXiv:1111.6097 [hep-ph]].

\bibitem{Degrande:2011ua} 
  C.~Degrande, C.~Duhr, B.~Fuks, D.~Grellscheid, O.~Mattelaer and T.~Reiter,
  Comput.\ Phys.\ Commun.\  {\bf 183}, 1201 (2012)
  [arXiv:1108.2040 [hep-ph]].

\bibitem{Dermisek:2013cxa} 
  R.~Dermisek, J.~P.~Hall, E.~Lunghi and S.~Shin,
  JHEP {\bf 1404}, 140 (2014)
  [arXiv:1311.7208 [hep-ph]].

\end{thebibliography}
\end{document}